\documentclass[twoside,leqno,twocolumn]{article}

\usepackage[letterpaper]{geometry}

\usepackage{ltexpprt}
\usepackage{lipsum,adjustbox}

\usepackage{caption}
\usepackage{tabularx}
\usepackage{makecell}
\usepackage{graphicx}
\usepackage{tikz}
\usepackage{lipsum}
\usepackage{xcolor}
\usepackage{amsopn}
\usepackage{amsfonts}

\usepackage{stackrel}
\usepackage{todonotes}
\usepackage[hypertexnames=false, draft]{hyperref}
\usepackage{mathtools}

\usepackage{algorithm}
\usepackage[noend]{algpseudocode}

\usepackage{amsmath}
\usepackage{amssymb}
\usepackage[capitalise]{cleveref}
\usepackage{subcaption}

\usepackage{flushend}

\makeatletter
\renewcommand{\ALG@name}{\sc Algorithm}
\makeatother

\usetikzlibrary{positioning, shapes.geometric}
\usetikzlibrary{shadows}
\usetikzlibrary{positioning,arrows,shapes.arrows}

\usepackage{xspace}
\newcommand{\probsty}[1]{\textsc{#1}\xspace}
\newcommand{\algsty}[1]{\texttt{#1}\xspace}

\usepackage{xargs}
\usepackage{xifthen}
\usepackage{framed}
\usetikzlibrary{calc}

\newenvironment{tightcenter}
 {\parskip=0pt\par\nopagebreak\centering}
 {\par\noindent\ignorespacesafterend}

\usepackage{ctable}
\newlength{\RoundedBoxWidth}
\newsavebox{\GrayRoundedBox}
\newenvironment{GrayBox}[1]%
{\setlength{\RoundedBoxWidth}{\linewidth-4.5ex}
\def\boxheading{#1}
\begin{lrbox}{\GrayRoundedBox}
\begin{minipage}{\RoundedBoxWidth}%
}{%
\end{minipage}
\end{lrbox}%
\begin{tightcenter}%
\begin{tikzpicture}%
\node(Text)[draw=black!20,fill=white,rounded corners,%
inner sep=2ex,text width=\RoundedBoxWidth]%
{\usebox{\GrayRoundedBox}};
\coordinate(x) at (current bounding box.north west);
\node [draw=white,rectangle,inner sep=3pt,anchor=north west,fill=white]
at ($(x)+(10.5pt,.75em)$) {\boxheading};
\end{tikzpicture}
\end{tightcenter}\vspace{0pt}%
\ignorespacesafterend
}

\newenvironment{problem}[1]{\noindent\ignorespaces%
\FrameSep=6pt%
\parindent=0pt%
\vspace*{.5em}
\begin{GrayBox}{\textsc{#1}}%
\newcommand\Input{Input:}%
\newcommand\Prob{Output:}%
\begin{tabular*}{\columnwidth}{@{\hspace{.25em}} >{\itshape} p{1.1cm} p{0.8\columnwidth} @{}}%
}{
\end{tabular*}%
\end{GrayBox}%
\vspace*{-.5em}
\ignorespacesafterend
}

\colorlet{leftclique}{blue!40!green!90!gray}
\colorlet{midclique}{red!60!blue}
\colorlet{rightcliquetop}{red!10!blue!70}
\colorlet{rightcliquebot}{red!70!yellow!50}

\newtheorem{reduction}{Rule}
\def\WCD{WCD\xspace}
\def\WCDfull{\probsty{Weighted Clique Decomposition}}
\def\AEWCDfull{\probsty{Annotated EWCD}}
\def\AEWCD{AEWCD\xspace}
\def\EWCD{EWCD\xspace}
\def\EWCDfull{\probsty{Exact Weighted Clique Decomposition}}
\def\WECP{WECP\xspace}
\def\WECPfull{\probsty{Weighted Edge Clique Partition}}

\def\WBSDDWfull{\probsty{Binary Symmetric Weighted Decomposition with Diagonal Wildcards}}
\def\WBSDDWprob{\probsty{BSWD-DW}}

\def\WBSDDW{BSWD-DW\xspace}
\def\BSWDDW{BSWD-DW\xspace}

\def\FillNB{\algsty{FillNonBasis}}
\def\Comp{\algsty{iWCompatible}}
\def\InferCWLp{\algsty{InferCliqWts-LP}}
\def\InferCWIp{\algsty{InferCliqWts-IP}}
\def\AlgLp{\algsty{CliqueDecomp-LP}}
\def\AlgIp{\algsty{CliqueDecomp-IP}}
\def\UpdWsIp{\algsty{UpdateWs}}

\def\for{\algsty{for}\xspace}
\def\while{\algsty{while}\xspace}
\def\feldalg{\algsty{wecp}}
\def\ipalg{\algsty{ipart}}
\def\lpalg{\algsty{lp}}

\def\cw{\gamma}

\def\es{\stackrel{\star}{=}}
\def\basis{\widetilde{B}}
\def\basisA{\widetilde{A}}
\def\basisE{E(\basis)}
\def\R{\mathbb{R}}
\def\bin{\left\{ 0,1 \right\}}

\def\weightlist{\textbf{W}}

\def\Sets{\mathcal{S}}	% the k value input to the algs
\def\C{\mathcal{C}}	% the k value input to the algs

\begin{document}

\title{\Large Parameterized algorithms for identifying gene co-expression modules via weighted clique decomposition\thanks{This work was supported by the NIH R01 HG010067 and the Gordon \& Betty Moore Foundation under awards GBMF4552 and GBMF4560.}}
\author{Madison Cooley\thanks{University of Utah, \texttt{mcooley@cs.utah.edu}}
\and Casey S. Greene\thanks{University of Colorado School of Medicine, \hfill \break
	\texttt{greenescientist@gmail.com}}
\and Davis Issac\thanks{Hasso Plattner Institute, \texttt{davis.issac@hpi.de}}
\and Milton Pividori\thanks{University of Pennsylvania, \hfill \break
	\texttt{milton.pividori@pennmedicine.upenn.edu}}
\and Blair D. Sullivan\thanks{University of Utah, \texttt{sullivan@cs.utah.edu}}}

\date{\today}

\maketitle

% Copyright Statement
% When submitting your final paper to a SIAM proceedings, it is requested that you include
% the appropriate copyright in the footer of the paper.  The copyright added should be
% consistent with the copyright selected on the copyright form submitted with the paper.
% Please note that "20XX" should be changed to the year of the meeting.

% Default Copyright Statement
\fancyfoot[R]{\scriptsize{Copyright \textcopyright\ 2021 by SIAM\\
Unauthorized reproduction of this article is prohibited}}

% Depending on which copyright you agree to when you sign the copyright form, the copyright
% can be changed to one of the following after commenting out the default copyright statement
% above.

%\fancyfoot[R]{\scriptsize{Copyright \textcopyright\ 20XX\\
%Copyright for this paper is retained by authors}}

%\fancyfoot[R]{\scriptsize{Copyright \textcopyright\ 20XX\\
%Copyright retained by principal author's organization}}

%\pagenumbering{arabic}
%\setcounter{page}{1}%Leave this line commented out.

\begin{abstract} \small\baselineskip=9pt

We present a new combinatorial model for identifying regulatory modules in gene co-expression data using a decomposition into weighted cliques. To capture complex interaction effects, we generalize 
the previously-studied weighted edge clique partition problem. As a first step, we restrict ourselves to the noise-free setting, and show that the problem is fixed parameter tractable when parameterized by the number of modules (cliques). We present two new algorithms for finding these decompositions, using linear programming and integer partitioning to determine the clique weights. Further, we implement these algorithms in Python and test them on a biologically-inspired synthetic corpus generated using real-world data from transcription factors and a latent variable analysis of co-expression in varying cell types.
\end{abstract}

\section{Introduction}
Biomedical research has recently seen a burgeoning of methods that incorporate network analysis to improve understanding and prediction of complex phenotypes~\cite{Greene2015}.
These approaches leverage information encoded in the interactions of proteins or genes, which are naturally modeled as graphs.
Further, there has been an explosion of available data including large gene expression compendia~\cite{ColladoTorres2017,Lachmann2018} and protein-protein interaction maps~\cite{Venkatesan2009}\looseness-1.

A core problem in this area has always been identifying groups of co-acting 
genes/proteins, which often manifest as a clique or dense subgraph in the resulting network.
In this work, we consider the specific setting of identifying gene co-expression modules (or pathways) from large datasets, with a downstream objective of aiding the development of new therapies for human disease.

There is substantial evidence that drugs with genetic support are more likely to progress through the drug development pipeline~\cite{Nelson2015-ag}.
Prior work has shown that approaches that consider genes’ roles in biological networks can be robust to gene mapping noise~\cite{Leeuw2015}, which might suggest alternative treatment avenues when a directly associated gene cannot be targeted.

Unfortunately, the membership of genes in modules and the relative strength of effect a module has on co-expression of its constituents are not directly observable. In gene co-expression analysis, what we are able to obtain is pairwise correlations for all genes in the organism~\cite{Mercatelli2020}. Existing approaches rely on machine-learning to identify clusters in these data sets~\cite{Leeuw2015,Mao2019}; here, we propose a new combinatorial model for the problem\looseness-1.

By modeling the observed gene expression data as a projection of a weighted bipartite graph representing gene-module membership and strength of expression for each module, we can represent the problem as a decomposition of the co-expression network into a collection of (potentially overlapping) weighted cliques (we call this \WCDfull).

While the resulting problem is naturally NP-hard, we demonstrate that techniques from parameterized algorithms enable efficient approaches when the number of modules is small. We present two parameterized algorithms for solving this problem\footnote{one of which restricts to integral edge weights}; both run in polynomial time in the network size, but have exponential dependence on the number of modules. As a first step towards practicality, we implement these methods\footnote{code is available at https://github.com/TheoryInPractice/cricca} 
and provide preliminary experimental results on biologically-inspired synthetic networks with ground-truth modules derived from data on gene transcription factors and gene co-expression modules identified using a machine-learning approach.

\section{Motivating Biological Problem}

Complex human traits and diseases are caused by an intricated molecular machinery that interacts with environmental factors.
For example, although asthma has some common features such as wheeze and shortness of breath, research suggests that this highly heterogeneous disease is comprised of several conditions~\cite{Wenzel2012}, such as childhood-onset asthma and adult-onset asthma, which present different prognosis and response to treatment, and also differ in their genetic risk factors~\cite{Pividori2019-bn}.
Genome-wide association studies (GWAS) are designed to improve our understanding of how genetic variation leads to phenotype by detecting genetic variants correlated with disease.
GWAS have prioritized causal molecular mechanisms that, when disturbed, confer disease susceptibility, and these findings were later translated to new treatments~\cite{Visscher2017-zd}.
Drug targets backed by the support of genetic associations are more likely to succeed through the process of clinical development~\cite{Nelson2015-ag}.
However, understanding the influence of genetic variation on disease pathophysiology towards the development of effective therapeutics is complex.
GWAS often reveal variants with small effect sizes that do not account for much of the risk of a disease~\cite{Tam2019}.
On the one hand, widespread gene pleiotropy (a gene affecting several unrelated phenotypes) and polygenic traits (a single trait affected by several
genes) reveal the highly interconnected nature of biomolecular networks~\cite{Moore2010,Cordell2009}.

Instead of looking at single gene-disease associations, methods that consider groups of genes that are functionally related (i.e., that belong to the same pathways) can be more robust to identify putative mechanisms that influence disease, and also provide alternative treatment avenues when directly associated genes are not druggable~\cite{Menche2015,Dozmorov2020}.
Large gene expression compendia such as recount2~\cite{ColladoTorres2017} or ARCHS4~\cite{Lachmann2018} provide unified resources with publicly available RNA-seq data on tens of thousands of samples.
Leveraging this massive amount of data, unsupervised network-based learning approaches~\cite{Mao2019,Taroni2019,Leeuw2015} can detect meaningful gene co-expression patterns: sets of genes whose expression is consistently modulated across the same tissues or cell types.
However, this is particularly challenging because the observed data is an aggregated and noisy projection of a highly complex transcriptional machinery: co-expressed genes can be controlled by the same regulatory program or module, but single genes can also play different roles in different modules expressed in distinct tissues or cell differentiation stages~\cite{Bush2016,Watanabe2019}.
For example, Marfan syndrome (MFS) is a rare genetic disorder caused by a mutation in gene \textit{FBN1}, which encodes a protein that forms elastic and nonelastic connective tissue~\cite{NHLBIMFS}.
However, MFS is characterized by abnormalities in bones, joints, eyes, heart, and blood vessels, suggesting that \textit{FBN1} is implicated in independent pathways across different tissues or cell types.
In other words, the membership of genes in modules and the relative strength of effect a module has on the co-expression of its constituents are not directly observable from gene expression data.

%Figure illustrating the problem and definitions

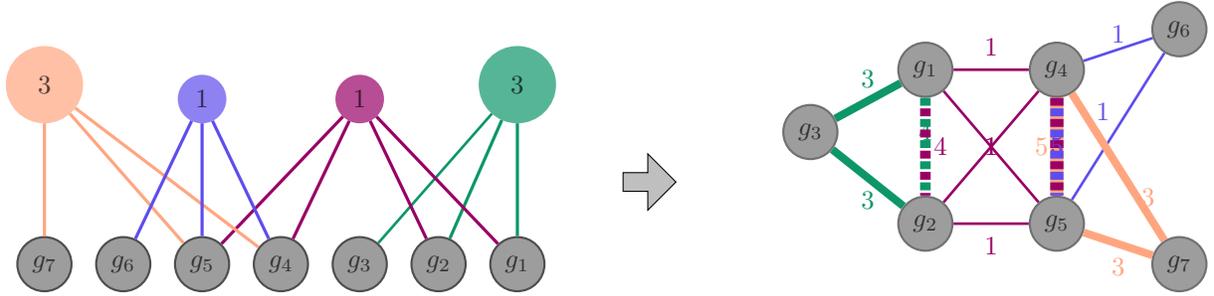
\begin {figure*}%[!hbtp]
\centering
\begin{tikzpicture}[scale=0.20] %[x={10.0pt},y={3.0pt}]
\tikzstyle{node} = [circle, fill=gray!90!black, draw, thick, opacity=0.7]
\tikzstyle{clique} = [circle, fill=gray!99!black, draw=black!80, thick, opacity=0.7,]
	\tikzstyle{edge} = [thick]
	\tikzstyle{dim} = [opacity=0.3]

	% ------Nodes
	%clique left
	\node (1) [node] {$g_1$};
	\node (2) [node, left=0.3cm of 1] {$g_2$};
	\node (3) [node, left=0.3cm of 2] {$g_3$};
	\node (4) [node, left=0.3cm of 3] {$g_4$};
	\node (5) [node, left=0.3cm of 4] {$g_5$};
	\node (6) [node, left=0.3cm of 5] {$g_6$};
	\node (7) [node, left=0.3cm of 6] {$g_7$};

	\node (8) [clique, leftclique, text=black, above=1.5cm of 1,
			%label={[shift={(-0.1,0.2)}]$w=3$},
			minimum size=1cm] {$3$};

	\node (9) [clique, midclique, text=black, above=1.5cm of 3,
			%label={[shift={(-0.1,0.2)}]$w=1$},
			] {$1$};

	\node (10)[clique, rightcliquetop, text=black, above=1.5cm of 5,
			%label={[shift={(-0.1,0.2)}]$w=1$},
			] {$1$};

	\node (11)[clique,  rightcliquebot, text=black, above=1.5cm of 7,
			%label={[shift={(-0.1,0.2)}]$w=3$},
			minimum size=1cm] {$3$};

	\draw[leftclique, line width=1.2pt] (1) edge node[swap] {} (8);
	\draw[leftclique, line width=1.2pt] (2) edge node[swap] {} (8);
	\draw[leftclique, line width=1.0pt] (3) edge node[swap] {} (8);

	\draw[midclique, line width=1.2pt] (1) edge node[swap] {} (9);
	\draw[midclique, line width=1.2pt] (2) edge node[swap] {} (9);
	\draw[midclique, line width=1.2pt] (4) edge node[swap] {} (9);
	\draw[midclique, line width=1.2pt] (5) edge node[swap] {} (9);

	\draw[rightcliquebot, line width=1.2pt] (7) edge node[swap] {} (11);
	\draw[rightcliquebot, line width=1.2pt] (5) edge node[swap] {} (11);
	\draw[rightcliquebot, line width=1.2pt] (4) edge node[swap] {} (11);

	\draw[rightcliquetop, line width=1.2pt] (6) edge node[swap] {} (10);
	\draw[rightcliquetop, line width=1.2pt] (5) edge node[swap] {} (10);
	\draw[rightcliquetop, line width=1.2pt] (4) edge node[swap] {} (10);

\node[single arrow, draw=black, fill=black!25, minimum height=2em,
		below right=0.8cm and 1.2cm of 8](arrow){};

\end{tikzpicture}% pic 1
\qquad % <----------------- SPACE BETWEEN PICTURES
\qquad
\begin{tikzpicture}[scale=0.50] %[x={10.0pt},y={3.0pt}]
\tikzstyle{node} = [circle, fill=gray!90!black, draw=black!80, thick, opacity=0.7]
	\tikzstyle{edge} = [thick]
	\tikzstyle{dim} = [opacity=0.3]

	% ------Nodes
	%clique left
	\node (1) [node] {$g_1$};
	\node (2) [node, below=1.3cm of 1] {$g_2$};
	\node (3) [node, below left=0.3cm and 1.0cm of 1] {$g_3$};

	%clique middle
	\node (4) [node, right=1.0cm of 1] {$g_4$};
	\node (5) [node, right=1.0cm of 2] {$g_5$};

	%clique right-top
	\node (6) [node, above right=0.01cm and 1.1cm of 4] {$g_6$};

	%clique right-bottom
	\node (7) [node, below right=0.01cm and 1.1cm of 5] {$g_7$};

	% -------Edges
	%clique left
	\draw[leftclique, line width=3.0pt] (1) edge node[swap, yshift=0.3cm] {3} (3);

	% shared edge
	\draw[leftclique, dash pattern= on 3pt off 5pt,
		 line width=4.0pt] (1) edge node[swap, yshift=0.2cm] {4} (2);
	\draw[midclique, dash pattern= on 3pt off 5pt,dash phase=4pt,
		 line width=4.0pt] (1) edge node[swap, xshift=0.2cm] {4} (2);
	\draw[leftclique, line width=3.0pt] (2) edge node[yshift=-0.3cm]{3} (3);

	%clique middle
	\draw[midclique, line width=1.0pt, minimum size=0.2cm] (4) edge node[swap, yshift=0.3cm] {1} (1);
	\draw[midclique, line width=1.0pt] (4) edge node[swap] {1} (2);
	\draw[midclique, line width=1.0pt] (2) edge node[swap,  yshift=-0.3cm] {1} (5);
	\draw[midclique, line width=1.0pt] (1) edge node{1} (5);

	%clique right top/bot
	% shared edge
	\draw[rightcliquebot,
		line width=5.0pt] (4) edge node[swap, xshift=-0.2cm] {5} (5);
	\draw[midclique, dash pattern= on 3pt off 5pt,
		line width=5.0pt] (4) edge node[swap] {5} (5);
	\draw[rightcliquetop, dash pattern= on 3pt off 5pt,dash phase=4pt,
		line width=5.0pt] (4) edge node[swap] {5} (5);

	\draw[rightcliquetop, line width=1.0pt] (4) edge node[yshift=0.2cm]{1} (6);
	\draw[rightcliquebot, line width=3.0pt] (5) edge node[swap, yshift=-0.3cm] {3} (7);
	\draw[rightcliquetop, line width=1.0pt] (5) edge node[xshift=-0.2cm,  yshift=0.2cm]{1} (6);
	\draw[rightcliquebot, line width=3.0pt] (4) edge node[yshift=-0.4cm, xshift=0.4cm]{3} (7);
\end{tikzpicture}% pic 2
\caption{A bipartite graph (left) of genes $g_1, \ldots g_7$ and modules (top, labelled with strength of expression) gives rise to a gene-gene interaction network (right) with edges weighted by the sum of the strengths of all modules that contain both endpoints (indicated by color coding).}
\label{fig:probmod}
\end{figure*}

\section{Problem Modeling}

We begin by observing that gene-module membership is naturally represented by a bipartite graph $B$, where each gene has an edge to all modules it participates in. Further, in order to capture the notion of varied effect-strength among modules, we associate a non-negative real-valued weight $w_i$ to each module $c_i$, since we are interested in sets of co-expressed genes. In other words, we assume that all pairs of genes that are common to module $i$ will be co-expressed with strength $w_i$; thus, the genes in each module will form a clique in the co-expression network. Further, we assume that modules interact with one another in a linear, additive manner. That is, the co-expression between genes $u$ and $v$ is the sum of the weights of all modules containing both $u$ and $v$. In a noise-free setting, this means that the gene-gene co-expression network is exactly a union of (potentially overlapping) cliques $m_1, \ldots m_k$ with associated weights $w_1, \ldots w_k$ so that the weight on $uv$ is exactly $\sum_{\{u,v\} \subseteq m_i} w_i$. It is important to note that not all valid solutions are interesting; specifically, one can always assign each pair of genes to its own clique of size 2, and get a valid solution. We rely on the principal of parsimony, and try to find an assignment which minimizes the number of modules in a valid solution. Realistically, the edge-weights will not satisfy exact equality, and we will need to consider an optimization variant of our problem which minimizes an objective function incorporating penalties for over/under-estimating the observed co-expressions\looseness-1.

To this end, we introduce a penalty function $\phi$ on the edges based on the discrepancy between the weight predicted by clique (module) membership and the original weight (observed co-expression value), then minimize $\phi$ to determine an optimal solution. For example, a natural choice for $\phi$ might be the sum of the absolute value of the discrepancies on each edge. Formally, this leads to the following problem:

\begin{problem}{\WCDfull}
\Input & a graph $G = (V,E)$, a non-negative weight function $w_e$ on $E$, a penalty function $\phi$, and a positive integer $k$.\\
\Prob  & a set of at most $k$ cliques $C_1, \ldots C_k$ with weights $\cw_1, \ldots \cw_k\in\R^+ $ that define
$\cw_{uv} = \sum_{i:uv \in C_i} \cw_i$ for all $uv \in E$, such that $\phi(\{(w_e, \cw_e):e\in E\})$ is minimized.\\
\end{problem}

In the remainder of this paper, we restrict our attention to the setting where equality can be achieved (as one might expect in synthetic data); further discussion of ideas for addressing the optimization variant is deferred to the future work section.
For convenience, we define a decision version of \WCD for this setting (this is equivalent to having a penalty function which is zero for matching the weight on an edge and infinite for any discrepancy):

\begin{problem}{\EWCDfull}
\Input & a graph $G = (V,E)$, a non-negative weight function $w_e$ for $e\in E$, and a positive integer $k$.\\
\Prob & a set of at most $k$ cliques $C_1, \ldots C_k$ with weights $\cw_1, \ldots \cw_k\in \R^+$ such that
$w_{uv} = \sum_{i:{uv} \in C_i} \cw_i$ for all $uv \in E$ (if one exists, otherwise output NO).\\
\end{problem}

If the clique weights are constrained to be integers then the problem becomes a generalization of the NP-hard problem \probsty{Edge Clique Partition}~\cite{ma1988complexity,fleischer2010K4Planar}. The NP-hardness of the fractional-clique-weight version also follows easily from the reduction in \cite{ma1988complexity}. For completeness, we give the proof in \cref{app:NPhard}.

\subsection{Annotated and Matrix Formulations}

We will work with the following more general version of \EWCD in our algorithms, where some of the vertices are annotated with vertex weights.
\begin{problem}{\AEWCDfull}
\Input & a graph $G = (V,E)$, a non-negative weight function $w_e$ for $e\in E$, a special set of vertices $S\subseteq V$, a non-negative weight function $w_v$ for $v\in S$, and a positive integer $k$.\\
\Prob & a set of at most $k$ cliques $C_1, \ldots C_k$ with weights $\cw_1, \ldots \cw_k\in \R^+$ such that
$w_{uv} = \sum_{i:{uv} \in C_i} \cw_i$ for all $uv \in E$ and
$w_{v} = \sum_{i:{v} \in C_i} \cw_i$
for all $v \in S$
(if one exists, otherwise output NO).\\
\end{problem}

Note that \EWCD is the special case of \AEWCD when the set $S=\varnothing$.

We also introduce an equivalent matrix formulation of \AEWCD,
as our techniques are heavily based on linear algebraic properties.
For this we use matrices that allow wildcard entries denoted by $\star$.
For $a,b\in \R\cup \left\{ \star \right\}$, we say $a\es b$ if either $a=b$ or $a=\star$ or $b=\star$.
For matrices $A$ and $B$, we say $A\es B$ if $A_{ij}\es B_{ij}$ for each $i,j$.
We call the matrix problem as \WBSDDWfull. Note that \EWCD is the special case where all the diagonal entries are wildcards.
\begin{problem}{\WBSDDWprob}
\Input &
a symmetric matrix $A\in \left(\R^+_0\cup\{\star\}\right)^{n\times n}$ with wildcards appearing on a subset of diagonal entries, and a positive integer $k$\\
\Prob &
a matrix $B\in \bin^{n\times k} $ and a diagonal matrix $W\in (\R_0^+)^{k\times k}$ such that
${A\es BWB^T}$.
(if such (B,W) exist, otherwise output NO).\\
\end{problem}

\section{Parameterized Algorithms} 
\label{sec:parameterized_algorithms}

Parameterized algorithms are a method used to tackle NP-hard problems where, besides the input size $n$, we are given an additional parameter $k$, most often representing the solution size.
E.g., in our problem WCD, the parameter $k$ is the number of cliques.
An algorithm is said to be \emph{fixed parameter tractable} if the runtime is polynomial in the input size and exponential only in the parameter---often resulting in tractable algorithms when the parameter is much smaller compared to the input size.
One of the most effective tools in parameterized algorithms is \emph{kernelization}, which is essentially a preprocessing framework that reduces the input to an equivalent instance of the same problem whose size depends only on the parameter $k$.
The reduced instance is called a \emph{kernel}.
Sometimes, the reduction is not to the same problem itself but to a different related problem, in which case it is called a \emph{compression}.
For an extensive introduction to the topic, we refer to the book by Cygan et al.~\cite{Parabook}\looseness-1.
%end of  section parameterized_algorithms 

\section{Prior Work}
The \WCD problem with integer clique weights is a generalization of the \WECPfull problem which in turn generalizes \probsty{Edge Clique Partition} \cite{ma1988complexity})\looseness-1:

\begin{problem}{\WECPfull}
	\Input & a graph $G=(V,E)$, a weight function $w_e : E \rightarrow Z^+$ and a positive integer $k$.\\
\Prob & a set of at most $k$ cliques such that each edge appears in exactly as many cliques as its weight (if it exists, otherwise output NO).\\
\end{problem}

\WECPfull (\WECP) was introduced by Feldmann et al.~\cite{feldmann2020} last year. They gave a $4^k$-compression and a $2^{\mathcal{O}(k^{3/2}w^{1/2}\log(k/w))}+\mathcal O(n^2\log n)$ time algorithm for WECP, where $w$ is the maximum edge weight.
The compression is into a more general problem called \probsty{Annotated Weighted Edge Clique Partition} (AWECP) where some vertices also have input weights and these vertices are constrained to be in as many cliques as its weight in the output.
The authors worked with an equivalent matrix formulation for AWECP called \probsty{Binary Symmetric Decomposition with Diagonal Wildcards} (BSD-DW) where given a $n\times n$ symmetric matrix $A$ with wildcards (denoted by $\star$) in the diagonal, the task is to find a $n\times k$ binary matrix $B$ such that $BB^T \es A$ where
$\es$ denotes that the wildcards are considered equal to any number.
The algorithm of Feldmann et al.~\cite{feldmann2020} builds upon the linear algebraic techniques used by Chandran et al.~\cite{Chandran2016Biclique} for solving the \probsty{Biclique Partition} problem.
Our algorithms further build upon the techniques of \cite{feldmann2020}.
Note that one could encode the clique weights (in the integer weight case) into the \WECP problem by thinking of a clique of weight $w$ as $w$ identical unweighted cliques. This makes the parameter $k$ equal to the sum of clique weights, and hence the algorithms of Feldmann et al.~\cite{feldmann2020} are not sufficient for our application.

The unit-weighted case of \WECP called \probsty{Edge Clique Partition} (ECP) has been more well studied, especially from the parameterized point of view. It is known that ECP admits a $k^2$-kernel in polynomial time~\cite{MujuniRosamond2008Kernel}. The fastest FPT algorithm for ECP is the algorithm by Feldmann et al.~\cite{feldmann2020} which runs in $2^{\mathcal{O}(k^{3/2}\log k)}+\mathcal{O}(n^2\log n)$ for ECP.
There are faster algorithms for ECP in special graph classes, for instance a $2^{\mathcal O(\sqrt{k})}n^{\mathcal O(1)}$ time algorithm for planar graphs, $2^{dk}n^{\mathcal O(1)}$ time algorithm for graphs with degeneracy $d$, and a $2^{\mathcal O(k)}n^{\mathcal O(1)}$ time algorithm for $K_4$-free graphs~\cite{fleischer2010K4Planar}.
A closely related problem to ECP is the \probsty{Edge Clique Cover} problem. Here, each edge should be present in at least one clique but can be present in any number of cliques. This unrestricted covering version is much harder and is known to \emph{not} admit algorithms running faster than $2^{2^{o(k)}}n^{\mathcal O(1)}$~\cite{Cygan2016ECC}.

There are a few papers that study symmetric matrix factorization problems that are similar to the \probsty{Binary Symmetric Decomposition with diagonal Wildcards} (BSD-DW) problem,
defined by Feldmann et al.~\cite{feldmann2020}.
Recall that BSD-DW is equivalent to the AWECP problem.
Zhang et al.~\cite{zhang2013symmetricBMF} studied the objective of minimizing ${\|A-BB^T\|_2^2}$.
Their matrix model does not translate into the clique model as they do not have wildcards in the diagonal.
Chen et al.~\cite{chen2021Instahide} studied the objective of minimizing ${\|A-BB^T\|_0}$, but also without wildcards.
A matrix model that has diagonal wildcards was considered by Moutier et al.~\cite{offdiagonal} under the name Off-Diagonal Symmetric Non-negative Matrix Factorization, but they allow $B$ to be any non-negative matrix and not just binary\looseness-1.

The non-symmetric variants of these matrix problems known as \probsty{Binary Matrix Factorization}, have been receiving a lot of attention recently~\cite{Miettinen2020Recent, Fomin2019ApxSchemes, Fomin2020Parameterized, Ban2019Ptas, Kumar2019, Chandran2016Biclique}. Here the objective is to minimize $\|A-BC\|$, where
$A$ is an $m\times n$ input matrix, $B$ is an $m\times k$ output binary matrix and $C$ is a $k\times n$ output binary matrix.
For example, a constant approximation algorithm running in $2^{\mathcal O (k^2\log k)}(mn)^{\mathcal O(1)}$ is known~\cite{Kumar2019}.
In the graph-world the non-symmetric problems correspond to finding a partition of the edges of a bipartite graph into bicliques (complete bipartite graphs) instead of cliques~\cite{Chandran2016Biclique}.

\section{Algorithms}
We give two algorithms for \WBSDDW and hence also for the equivalent \AEWCD and the special case \EWCD.
Both algorithms will have a common framework similar to that of Feldmann et al.~\cite{feldmann2020}.
The algorithms will differ in the method in which the clique weights (represented by the diagonal matrix $W$) are inferred. One uses an LP based method while the other uses an integer partition dynamic program\looseness-1.

The first step in our pipeline is to preprocess disjoint cliques and cliques that overlap only on single vertices (and thus have no overlapping edges) out of each graph. The specifics of this process are outlined in Appendix~\ref{appendix:preprocessing}, but it essentially runs a modified breadth-first search algorithm.  
Similar to Feldmann et al.~\cite{feldmann2020}, the second step in our algorithms is to give a kernel.
The kernel follows the same reduction rules as in Feldmann et al.~\cite{feldmann2020} i.e. by reducing blocks of twin vertices.
The proof of correctness follows analogously, and we omit it here due to space constraints.
After the kernelization, we can assume that the number of vertices of $G$ (equivalently the number of rows of matrix $A$) is at most $4^k$\looseness-1.

\begin{theorem}
	\label{thm:kernel}
	\AEWCD $($\WBSDDWprob resp.$)$ has a kernel with at most $4^k$ vertices $(4^k$ rows resp.$)$ that can be found in $\mathcal O({n^3})$ time.
\end{theorem}

The third step is to run a clique decomposition algorithm on the kernelized \AEWCD instance to obtain the clique assignments for each vertex and clique weights.
Let $A$ be the input instance for \WBSDDW and let $G$ be the corresponding input instance to \AEWCD.
Both our algorithms use the basis-guessing principle used by Feldmann et al.~\cite{feldmann2020}, first introduced by
Chandran et al.~\cite{Chandran2016Biclique}.
The principle is that
once we correctly guess the entries of a row-basis of $B$,
then the remaining rows of $B$ can be filled iteratively without backtracking.
However, the technique does not carry over directly to the clique-weighted problem we have here.
We additionally need to infer the clique-weight matrix $W$, which poses some additional challenges.
Note that it is not feasible to guess the entries of the diagonals of $W$ as each entry could be as large as the largest element in $A$.
So once we have a guess for the basis, we also need to infer compatible values of $W$.
Since there could be multiple choices for compatible $W$, we are not guaranteed to hit the correct solution for $W$, likely producing some backtracking while filling the non-basis rows.
We tackle this by showing that if we guess a row-basis plus an additional $k$ rows (thus at most $2k$ rows) then the choice of $W$ does not matter.
If our guess for this $2k$ rows (we call it the \emph{pseudo-basis} of $B$) is correct, then we show that we can fill the other rows iteratively without any backtracking.
The intuition of why we need the additional $k$ rows is as follows: in the version without the $W$ matrix, once we fix the basis $\basis$ fo $B$, the matrix $\basisA$ given by the corresponding rows of $A$ is fixed. In particular the diagonal values $\basisA_{ii}$ (that could have been wildcards and hence not fixed apriori) are now fixed.
But with the matrix $W$, for different choices of $W$, we get different diagonal entries in $\basisA$.
We only need to add at most one more row to the pseudo-basis in order to fix one of these diagonal entries.
Thus we need at most $k$ additional rows in the pseudo-basis.

We guess the rows of the pseudo-basis on-demand i.e., we add a row as basis-row only if a compatible row cannot be found for it under the current inferred clique weights $W$ from the current basis matrix.
Every time we add a new row to the basis, we recompute the clique weights $W$.
The two algorithms that we present, differ in how they infer the clique weights for the current pseudo-basis.
The first algorithm uses a linear programming method while the second uses an integer partitioning dynamic programming method.
In our algorithms, we will often use partially filled matrices, i.e. some of the entries are allowed to have \emph{null} values.
If a row or matrix has all null values we call them null row and null matrix respectively.
\subsection{Clique Weight Recovery by Linear Programming}\label{section_algLp}

We use a linear program to infer the clique weights for the current pseudo-basis of the algorithm.
The pseudocode is given in {\InferCWLp} (Algorithm~\ref{alg:InferCWLp}).
Suppose $\basis$ is the current pseudo-basis matrix i.e. $\basis$ is an $n\times k$ matrix where the current pseudo-basis rows (at most $2k$) are filled by $0$'s and $1'$s,
and the other rows are null rows.
For each pair of distinct non-null rows $\basis_i$ and $\basis_j$ we add the constraint
$\basis_{i}^TW\basis_{j}=A_{ij}$ to the LP.
Also, for each $A_{ii}$ that is not a $\star$, we add the constraint $\basis_i^TW\basis_i=A_{ii}$.
Note that the variables of the LP are the diagonal entries $W_{11},W_{22},\cdots,W_{kk}$.
We also have non-negativity constraints $W_{11}\ge 0,\cdots,W_{kk}\ge 0$.
Any feasible solution to this LP gives us a set of clique weights compatible with the current pseudo-basis.
If the LP is infeasible, then we conclude that the current pseudo-basis guess is infeasible and proceed to the next guess.
Note that since we are only concerned about a feasible solution satisfying the constraints, we do not have an objective function for the LP.
We point out that solving this LP is rather efficient as the number of variables are $k$ and number of constraints are at most $4k^2$ and can be solved incrementally as we add constraints everytime a basis row is added.

\begin{algorithm}[!h]
\begin{algorithmic}[1]
	\For{$P \in \{0, 1 \}^{2k \times k}$}\label{line:LpFor}
\State{initialize $\basis$ to a $n \times k$ null matrix}
\State{$b, i \leftarrow 1$}
\While{$b \leq 2k $ }\label{line:LpWhile}
\State{$\basis_i \leftarrow P_b$}\label{line:LpBasisFill}
		\State{$b\leftarrow b+1$}
		\State{$W\leftarrow$ \InferCWLp($A$, $\widetilde{B}$)}
		\If{$W$ is not null matrix}
			\State{$(B, i) \leftarrow$ \FillNB(A, $\widetilde{B}$, $W$)}\label{line:LpFillNBCall}
			\If{$i=n+1$}
				\Return {($B$, $W$) }\label{line:LpYesReturn}
			\EndIf
		\Else
			$\;b \leftarrow 2k+1$ \Comment{\small{\it null $W$; break out of while}}
		\EndIf
	\EndWhile
\EndFor
\State \Return No \label{line:LpNoReturn}
\end{algorithmic}
\caption[]{\hspace*{-4.3pt}{.} \AlgLp}\label{algLppsuedo}
\label{alg:Lpmain}
\end{algorithm}

\begin{algorithm}[!h]
\begin{algorithmic}[1]
	\State \textbf{let} $\cw_1,\cdots,\cw_k \geq 0$ be variables of the LP
	\For {all pairs of non-null rows $\basis_i,\basis_j$ s.t. $A_{ij}\neq \star$}
	\State Add LP constraint $\sum_{1\le q\le k}\basis_{iq}\basis_{jq}\cw_q = A_{ij}$
	\EndFor
	\If {the LP is infeasible}
		\Return {the null matrix}
	\Else {}
		 \Return {the diagonal matrix given by $\cw_1,\ldots, \cw_k$}
	\EndIf
\end{algorithmic}
\caption[]{\hspace*{-4.3pt}{.} \InferCWLp($A$, $\widetilde{B}$)}\label{InferCliqWtsLp}
\label{alg:InferCWLp}
\end{algorithm}

\begin{algorithm}[!h]
\begin{algorithmic}[1]
\State $B\leftarrow\basis$
\While{$B$ has a null row}\label{line:fillnbwhile}
	\State \textbf{let} $B_i$ be the first null row
	\For{$v\in \left\{ 0,1 \right\}^k$}\label{line:fillnbfor}
	\If{\Comp($A$, $B$, $W$, $i$, $v$)}\label{line:compcall}
	\State $B_i\leftarrow v$ \label{line:BFill}
		\State \textbf{goto} line 2
	\EndIf
	\EndFor
	\State \Return $(B,i)$ \Comment{\small{\it there is no $(i,W)$-compatible $v$}}
\EndWhile
\State \Return $(B, n+1)$ \Comment{\small{\it B has no null row}}
\end{algorithmic}
\caption[]{\hspace*{-4.3pt}{.} \FillNB($A$, $\widetilde{B}$, $W$)}\label{FillNonBasis}
\label{alg:fillNB}
\end{algorithm}

\begin{algorithm}[!h]
\begin{algorithmic}[1]
	\For {each non-null row $B_j$}\label{line:compfor}
	\If {$v^TWB_j\neq A_{ij}$}\label{line:compmulti}
		\Return false
		\EndIf
	\EndFor
	\If {$v^TWv\not\es A_{ii}$}
		\Return false
		\EndIf
	\State \Return true

\end{algorithmic}
\caption[]{\hspace*{-4.3pt}{.} \Comp($A$, $\widetilde{B}$, $W$, $i$, $v$)}\label{iWcompatible}
\label{alg:comp}
\end{algorithm}

Once we have inferred a $W$ compatible with the current pseudo-basis, we then try to fill the remaining rows (we call them non-basis rows) one by one in \FillNB.
We say that a vector $v\in \bin^k$ is $(i,W)$ compatible with row $B_j$ if $v^TWB_j=A_{ij}$.
We say that $v$ is $(i,W)$ compatible with matrix $B$ if it is $(i,W)$-compatible with
each non-null row $B_j$,
and $v^TWv\es A_{ii}$.
We say $B$ and $W$ are compatible with each other if for each pair of non-null rows $B_i$ and $B_j$, we have $B_i^TWB_j\es A_{ij}$.
We keep filling the rows $B_i$ of $B$ one-by-one with $(i,W)$-compatible rows until either $B$ is completely filled or there is an $i$ such that there is no $(i,W)$-compatible vector in $\bin^k$.
In the former case we show that $(B,W)$ gives a solution, and in the latter case we proceed on to take row $i$ into the pseudo-basis row.
Note that when we take a new row into the pseudo-basis row we throw away all the non-basis rows and make them null rows again.
We will show that we only need to take up to $2k$ rows into the pseudo-basis for the algorithm to correctly find a solution.
\subsubsection{Algorithm Correctness}\label{section:Lpcorrectness}
\begin{theorem} \label{thm:Lp}
	{\AlgLp} (Algorithm~\ref{alg:Lpmain})
	correctly	solves the \BSWDDW problem, and hence also correctly solves \AEWCD and \EWCD, in time
	$\mathcal O(4^{k^2}k^2(32^k+k^3 L))$,
	where $L$ is the number of bits required for input representation.
\end{theorem}

First we prove in the following lemma that if {\AlgLp} outputs Yes, i.e. if it outputs through line~\ref{line:LpYesReturn}, then the matrices $B$ and $W$ output indeed satisfy that $A\es BWB^T$.
The proof follows because we checked for $(i,W)$-compatibility whenever we filled $B_i$.
The full proof of the Lemma can be found in \cref{section:missingLP}.
\begin{lemma}
If {\AlgLp} returns through line~\ref{line:LpYesReturn}, then the matrices $B$ and $W$ output satisfy that $A\es BWB^T$.
\label{lem:YesReturn}
\end{lemma}

\cref{lem:YesReturn} immediately implies that if the instance is a No-instance then the algorithm does not output through line~\ref{line:LpYesReturn}.
Since the only other possibility for output is through \cref{line:LpNoReturn}, which outputs No,
we can conclude that for a No-instance we correctly output No.
The following arguments are therefore related to the correctness of Yes instances.

For arguing the correctness in the Yes case, we fix a valid solution $(B^*,W^*)$ of the instance.
If the output occurs through \cref{line:LpYesReturn}, then by \cref{lem:YesReturn}, we are done.
So for the sake of contradiction assume that the output does not occur through \cref{line:LpYesReturn}.
For $I\subseteq [n]$, we define $B^*_I$ as the $n\times k$ matrix whose $i$-th row is equal to $B^*_i$ for all $i\in I$ and the other rows are null rows.
The following lemma follows because we iterate over all possible values of pattern matrix $P$.
\begin{lemma}
	Let $I\subseteq [n]$ be such that $|I|\le 2k-1$.
	If $\basis$ is equal to $B^*_I$ at some point in the algorithm,
and if \FillNB$(A,\basis=B^*_I,W)$ called in \cref{line:LpFillNBCall} returns $i\le n$,
then $\basis$ is equal to $B^*_{I\cup\left\{ i \right\}}$ at some point in the algorithm.
\label{lem:Bconsistent}
\end{lemma}

So, if we start with $I=\varnothing$, and repeatedly apply \cref{lem:Bconsistent},
then at some point in the algorithm, we have
$\basis=B^*_I$ such that $|I|=2k$.
Towards this, we define the matrix $\basisE$
formed by the rows that are element-wise products of pairs of non-null rows in $\basis$.
More precisely:
\begin{Definition}
	\label{def:basisE}
$\basisE$ is the matrix containing rows $\basis_i\odot \basis_j$ for each pair $i,j$ (not necessarily distinct) such that $A_{ij}\neq \star$.
Here $\odot$ denotes element-wise product.
\end{Definition}

The above definition means that $\basisE$ is the coefficient matrix of the LP that the algorithm would construct in the call \InferCWLp$(A,\basis)$.
\begin{Definition}[Pseudo-rank]
The pseudo-rank of $\basis$ is defined as the sum of ranks of $\basis$ and $\basisE$, where
by rank of $\basis$ we mean the rank of the matrix formed by the non-null rows of $\basis$.
\end{Definition}
Since the number of columns in $\basis$ and $\basisE$ are each $k$, we have the following lemma.
\begin{lemma}
	The pseudo-rank of $\basis$ is at most $2k$.
	\label{lem:pseudorank}
\end{lemma}
We say that a vector $v\in \bin^k$ \emph{$i$-extends} $\basis$ if $\basis_i$ is currently a null row, and adding $v$ as $\basis_i$ increases the pseudo-rank of $\basis$.

\begin{lemma}
	If \FillNB$(A,\basis=B^*_I,W)$ called on \cref{line:LpFillNBCall} returns $i\le n$, then $B^*_i$ $i$-extends $B^*_I$.
	\label{lem:iextend}
\end{lemma}
\begin{proof}
	Suppose for the sake of contradiction that $B^*_i$ does not $i$-extend $B^*_I$.
	This means that $B^*_i$ is linearly dependent on the non-null rows of $B^*_I$ and
	each $B^*_i\odot B^*_j$ for $j\in I$ is linearly dependent on the rows of $\basisE$.
	Also, if $A_{ii}\neq \star$ then $B^*_i\odot B^*_i$ is linearly dependent on the rows of $\basisE$.
	Now, consider each non-null row $B_j$ of the matrix $B$ when \Comp($A,B,W,i,v$) was called in \cref{line:compcall} in \FillNB.
We prove that $B^{*T}_iWB_j=A_{ij}$ and that $B^{*T}_iWB^*_i\es A_{ii}$.
This then implies that $B^*_i$ is $(i,W)$-compatible with $B$ and hence {\FillNB} could not have returned $i$,
giving a contradiction.

First consider the case when $B_j$ is a pseudo-basis row, i.e. $j\in I$.
Since $B^*_i$ does not $i$-extend $B^*_I$,
we know $B^*_i\odot B^*_j$ is linearly dependent on the rows of $\basisE$.
This means that adding $B^*_i$ as $\basis_i$ would not add any linearly independent equality constraints to the LP system that solves for $W$.
So, either the LP becomes infeasible or all the solutions to the LP still remain solutions.
But the LP is not infeasible as the diagonal elements of $W^*$ gives a feasible solution to the LP.
Thus, the current $W$ remains a feasible solution even after the addition of $B^*_i$ the row $\basis_i$.
Hence, $B^{*T}_iWB^*_j=A_{ij}$.

Now, consider the case when $B_j$ is not a pseudo-basis row, i.e. $j\notin I$.
In other words $\basis_j$ is a null row and $B_j$ was added
in {\FillNB}.
Since $B^*_i$ does not $i$-extend $B^*_I$,
we know that $B^*_i$ is linearly dependent on the non-null rows of $B^*_I$.
In other words, $B^*_i=\sum_{\ell\in I}\lambda_{\ell}B^*_{\ell}$ where each $\lambda_{\ell}\in \R$.
Then,
\begin{align}
	B^{*T}_iWB_j	&=\sum_{\ell\in I}\lambda_{\ell}B^{*T}_{\ell}WB_j\\
	&=\sum_{\ell\in I}\lambda_{\ell}A_{\ell j}\label{eq:BWBeA}\\
	&=A_{ij}\label{eq:Aij}
\end{align}
where \cref{eq:BWBeA} is because $B_j$ could have been selected for row $j$ only if it was $(j,W)$-compatible with $B^*_I$,
and \cref{eq:Aij} follows by using that $W^*B^*$ is a linear map from $B^*$ to $A$ and hence the linear dependencies in $B^*$ are preserved in $A$.
More precisely,
\begin{align*}
\sum_{\ell\in I}\lambda_{\ell}A_{\ell j}  &=
\sum_{\ell\in I}\lambda_{\ell}B^{*T}_{\ell}W^*B^*_j\\
&= B^*_iW^*B^*_j\\
&=A_{ij}
\end{align*}
Note that we have here crucially used $\ell\neq j$ (as $j\notin I$) and $j\neq i$ to say $=$ and not just $\es$.
This is the reason we required a separate argument for $j\in I$.
The argument for $B^{*T}_iWB^*_i\es A_{ii}$ follows the same argument as in the case of $j\in I$ by observing that if $A_{ii}\neq \star$ then $B^*_i\odot B^*_i$ is linearly dependent on the rows of $\basisE$.
\end{proof}
Now starting with $I=\varnothing$, and applying Lemmas~\ref{lem:Bconsistent} and \ref{lem:iextend} repeatedly, we have that at some point in the algorithm, $\basis$ is equal to some $B^*_I$ such that the pseudo-rank of $B^*_I$ is $2k$.
At this point, the {\FillNB} call at \cref{line:LpFillNBCall} should return $i=n+1$ because if it returned $i\le n$, then adding $B^*_i$ to $\basis$ would make the pseudo-rank of $\basis$ equal to $2k+1$, a contradiction to \cref{lem:pseudorank}.
Hence, the algorithm outputs through \cref{line:LpYesReturn}.
This concludes the correctness of the algorithm.
We defer the runtime analysis to \cref{section:Lpruntime}.

\subsection{Clique Weight Recovery by Integer Partitioning}
\def\barX{\overline{\textbf{X}}}
We give an algorithm for inferring the current pseudo-basis's clique weights by solving an integer partitioning dynamic program. The pseudocode is given in {\InferCWIp} (Algorithm~\ref{alg:InferCWIp}).
Similar to \ref{section_algLp}, consider $\basis$, the current pseudo-basis matrix.
Additionally, a list $\weightlist$ is maintained, containing partially filled diagonal weight matrices 
each of which are \emph{compatible} with the current pseudo-basis matrix $\basis$.
Here compatibility is defined as follows.
For a matrix $B$, first define its \emph{relevant indices}, denoted by $R(B)$, as the set of all $r\in [k]$ such that there exist non-null rows $B_i,B_j$ such that $B_{ir}B_{jr}=1$ and $A_{ij}\neq \star$.
For a diagonal matrix $W$ we define $F(W)$ as the set of all $i\in [k]$ such that $W_{ii}$ is not null.
We say that $\basis$ and $W$ are compatible if
$F(W)=R(\basis)$ and for each pair of non-null rows $\basis_i,\basis_j$ such that $A_{ij}\neq \star$,
it is true that $\sum_{r\in R(\basis)}\basis_{ir}W_{rr}\basis_{jr}= A_{ij}$.
We maintain in $\textbf{W}$, all the possible fillings of indices $R(B)$ of the diagonal vector of clique-weight matrix $W$ such that $W$ is compatible with $\basis$\looseness-1.

{\InferCWIp} is given as input the current $\basis$ after initially inserting $P_b$ at \cref{line:IpBasisFill} in {\AlgIp}. Thus,
$\basis_i$ is the potential basis row we are considering.
At the start of the function call, we know that
each non-null rows $\basis_j$ is $(j,W)$-compatible with each non-null row $\basis_{j'}$ for $j,j'\neq i$. 
If $\basis_i$ is not $(i,W)$-compatible, this implies that either there are null weights in $W$ in relevant positions, or the current basis $P$ being considered is not the correct one.
 Let $\basis_j$ be a non-null row such that $\basis_i$ is not $(i,W)$-compatible with $\basis_j$.
 We define $X\subseteq [k]$ as the indices of null positions in the
diagonal of $W$,
$\barX= [k]\setminus X$,
and $P\subseteq [k]$ as the set of positions in $\basis_i\odot\basis_j$ having $1$ values.
The sum $t=\sum_{l\in P\cap \barX}^k \basis_{il} \basis_{jl} W_{ll}$ is the sum of all previously fixed clique-weights that contribute to $A_{ij}$.
The difference $s=t-A_{ij}$ has to be contributed by $P\cap X$.
The {\UpdWsIp} function finds all possible ways to sum to $s$ using $|P\cap  X|$ number of non-negative integers via a dynamic program. This is an integer partitioning problem and is a simple variant of the common change-making problem.
Then for each such combination a new $W$-matrix is created by inserting the combination in the indices $P\cap X$.
{\UpdWsIp} returns the list of all such $W$-matrices created.
Note that $s$ could be negative in which case {\UpdWsIp} returns an empty list.

% Main IP alg
\begin{algorithm}[!h]
\begin{algorithmic}[1]
\For{$P \in \{0, 1 \}^{2k \times k}$}\label{line:IpFor}
\State{$\basis \leftarrow$ $n \times k$ null matrix}
\State{$b, i \leftarrow 1$}
\State{$\weightlist \leftarrow$ $\{$null matrix$\}$}
\While{$b \leq 2k $ }\label{line:IpWhile}
	\State{$\basis_i \leftarrow P_b$}\label{line:IpBasisFill}
	\State{$b\leftarrow b+1$}
	\State{$\textbf{S}$ $\leftarrow$ \InferCWIp($A$, $\basis$, $\weightlist$, $i$)}\label{line:IpInferCall}
	\If{$\textbf{S}$ is not empty}
		\State{$\weightlist \leftarrow \textbf{S}$}
		\State{$(B, i) \leftarrow$ \FillNB(A, $\basis$, $W[0]$)}\label{line:IpFillNBCall}
		\If{$i=n+1$}
		\Return {($B$, $W[0]$) }\label{line:IpYesReturn}
		\EndIf
	\Else{}
		\State{$b \leftarrow 2k+1$}
	\EndIf
\EndWhile
\EndFor
\State \Return No \label{line:IpNoReturn}
\end{algorithmic}
\caption[]{\hspace*{-4.3pt}{.} \AlgIp}\label{algIppsuedo}
\label{alg:Ipmain}
\end{algorithm}

% Clique weight finding IP routine
\begin{algorithm}[!h]
\begin{algorithmic}[1]
\State{$\textbf{S} \leftarrow [ \hspace{1.0pt} ]$}
\For{$l\leftarrow 1$ to $|\weightlist|$} \label{line:IpWloop1}

\State{$\textbf{X} \leftarrow$ \{$x\in [k]$ $|$ $W[l]_{xx}$ is null\}}\label{line:IpInferInitX}
\State{$\barX \leftarrow$ $[k]\setminus X$}\label{line:IpInferInitXbar}

	\State{$\textbf{temp} \leftarrow$ empty queue}
	\State{$\textbf{temp}$.push($\weightlist[l]$)}
	\For {each non-null row $\basis_j$ s.t. $A_{ij}\neq \star$} \label{line:IpCWbasis}
		\State{$ iters \leftarrow |\textbf{temp}| $}
	\State{$\textbf{P} \leftarrow$ \{$p\in [k]$ $|$ $\basis_{ip}\basis_{jp}=1$\}} \label{line:IpRelevant}
		\vspace*{2.0pt}
		\For{$q\leftarrow 1$ to $iters$} \label{line:IpCheckWs}
			\State{$S' \leftarrow \textbf{temp}$.pop()}
			\vspace*{2.0pt}
			\State{$s \leftarrow A_{ij} -
			\sum_{f \in \textbf{P}\cap \barX} S'_{ff}$} \label{line:Ipsumto}
			\vspace*{3.0pt}
			\State{$\textbf{T} \leftarrow$
			\UpdWsIp($S'$,$\textbf{P}\cap\textbf{X}$,
				$s$)}
			\State{$\textbf{temp}$.push($\textbf{T}$)}
			\If{$\textbf{temp}$ is empty}
			 goto \cref{line:SpushTemp}
			\EndIf
		\EndFor
	\EndFor
	\State{$\textbf{S}$.push($\textbf{temp}$)}\label{line:SpushTemp}
\EndFor
\State \Return {$\textbf{S}$} \label{line:IpCWreturn}
\end{algorithmic}
\caption[]{\hspace*{-4.3pt}{.} \InferCWIp($A$, $\basis$, $\weightlist$, $i$)}\label{InferCliqWtsIp}
\label{alg:InferCWIp}
\end{algorithm}

% DP routine
\begin{algorithm}[!h]
\begin{algorithmic}[1]
\State{$\textbf{V} \leftarrow$  [ \vspace{1.0pt} ]}
\State{$\textbf{Y} \leftarrow$ all partitions of $s$ into $|I|$ non-negative integers}
\For{each $parts$ in \textbf{Y}}\label{line:ForLoopParts}
\State{$y \leftarrow 0$}
\State{$C \leftarrow$ $W$}\label{line:assignCW}
\For{each $i \in \textbf{I}$}\label{line:InnerForLoopUpdateWs}
\State{$C_{ii} \leftarrow $ $parts[y]$ }
	\State{$y \leftarrow y+1$}
\EndFor
\vspace*{2.0pt}
\State{$\textbf{V}$.push($C$)}
\EndFor
\State \Return {$\textbf{V}$}
\end{algorithmic}
\caption[]{\hspace*{-4.3pt}{.} \UpdWsIp($W$, $\textbf{I}$, $s$)}\label{UpdateWeightsIp}
\label{alg:UpdateWsIp}
\end{algorithm}

\subsubsection{Algorithm Correctness}

\begin{theorem} \label{thm:Ip}
	{\AlgIp} (Algorithm~\ref{alg:Ipmain})
	correctly solves the \BSWDDW problem, and hence also correctly solves \AEWCD and \EWCD, in time 
$\mathcal O(4^{k^2} 32^kw^kk)$
where $w$ is the maximum entry of $A$. 
\end{theorem}
The proof of {\AlgIp}'s correctness is similar to the proof of {\AlgLp} in \cref{section:Lpcorrectness}. We describe the differences in Appendix~\ref{appendix:IP-correct} and defer the runtime analysis to \cref{section:Ipruntime}.\looseness-1

%%%%%%%%%%%%%% FIGURE PLACEMENT %%%%%%%%%%%%%%%%%%%%%%%%%%

%Figure including comparisons where we ran the Feldmann et al algorithm
\begin{figure}
	\centering
	\includegraphics[width=1.0\linewidth]{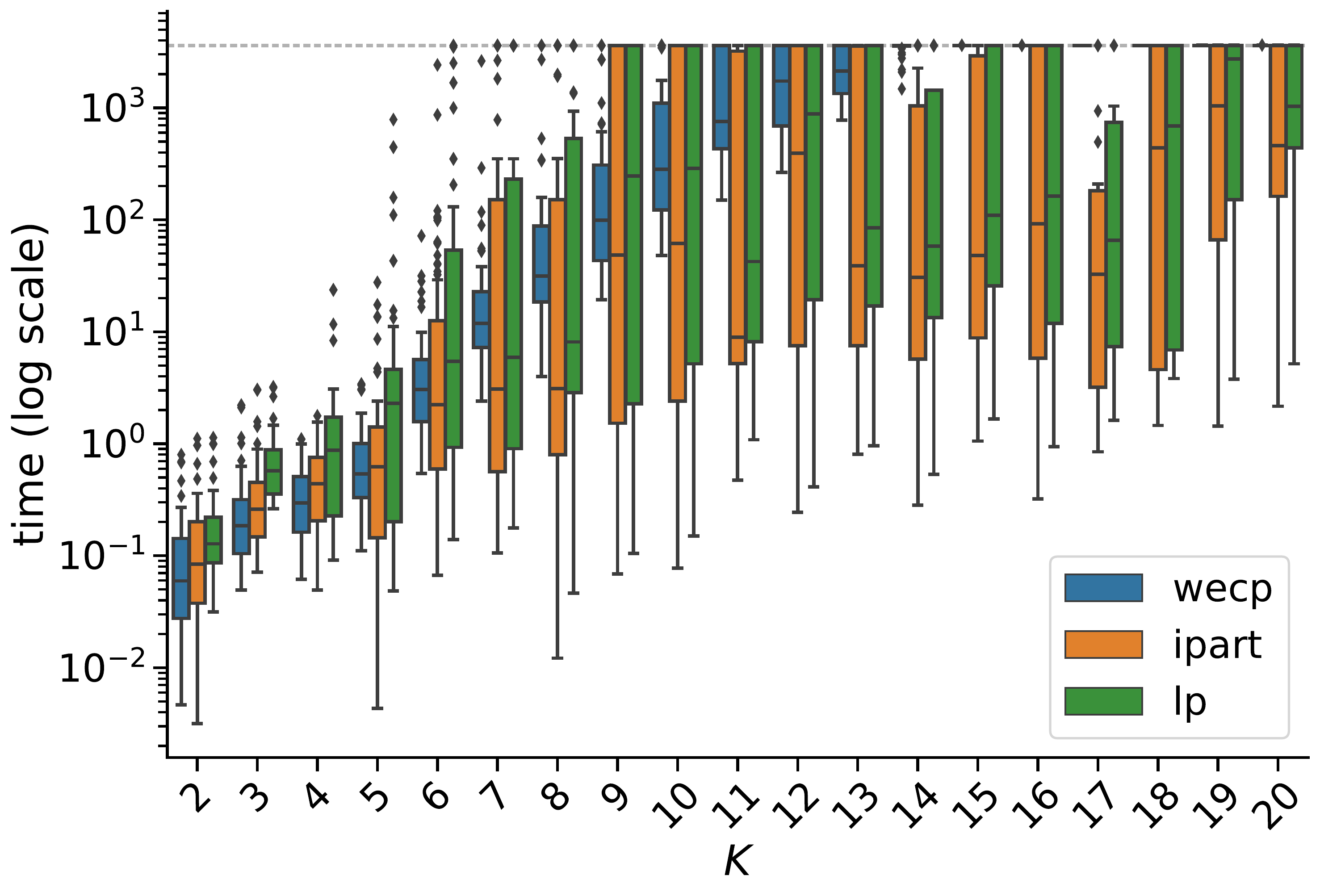}%
\caption{Log-scale plot showing distribution of total algorithm runtimes when binned by $K$ (the sum of the clique weights). All $K$ values shown in Figure~\ref{fig:orig-v-new-full} in Appendix~\ref{appendix:expresults}.\looseness-1}%
\label{fig:orig-v-new-2a}%
\end{figure}

%%%%%%%%%%%%%%%%%%%%%%%%%%%%%%%%%%%%%%%%%%%%%%%%%%%%%%%%%%

\section{Experimental Setup}
This section describes the synthetic corpora; hardware descriptions can be found in Appendix~\ref{appendix:hardware}.
We generate two sets of biologically-inspired synthetic graphs. The first dataset defines modules (cliques) using known relationships between transcription factors and genes; the second uses latent variables from a machine learning approach for analyzing co-expression data\looseness-1.

\subsection{TF-Dataset}\label{section:tfdata}
Our first dataset emulates the bipartite gene-module network by using known relationships between transcription factors (TFs) and genes~\cite{Essaghir2010}.
To generate a network with a ground truth of $k$ cliques, we randomly select $k$ TFs and form the network which is the union of all associated genes with edges between those that share at least one selected TF.

Since the relative strengths of effect on expression are unknown, we specify a desired maximum edge weight (see Appendix~\ref{appendix:dataSML}), and generate integral clique weights as described in Appendix~\ref{appendix:dataTF}.
A heavy-tailed distribution is chosen to mimic the view that modules have widely varying effects on gene co-expression, and a small minority likely have drastically higher impact than all others~\cite{Taroni2019}.

\subsection{LV-Dataset}

A similar approach is taken when generating the set of the latent variable-associated synthetic graphs. In this data~\cite{MultiPLIERLVs,Taroni2019}, each latent variable (LV) represents a set of genes that are co-expressed in the same cell types. A score for every gene in each LV indicates the strength of its association to the module. Further, some latent variables have been shown to align with prior knowledge of pathway associations~\cite{Taroni2019}. Our generator randomly selects $k$ latent variables, with 80\% drawn from those known to be aligned with pathways, and the remaining 20\% chosen uniformly from all LVs. For each LV, we only include genes with association scores above a threshold, determined as described in Appendix~\ref{appendix:dataLV}.

In contrast to the TF data, here the clique weights have a basis in the underlying data. For each LV, we compute the average associate score over all included genes then linearly transform this to control the maximum edge weight in the network (see Appendix~\ref{appendix:dataSML}).

\section{Results}
This section highlights the key outcomes of our preliminary experimental evaluation. We begin by highlighting
the effects of reparameterization, comparing the algorithm of~\cite{feldmann2020} (referred to as \feldalg)
to \AlgIp and \AlgLp (shortened to \ipalg and \lpalg for consistency with figure legends).

\subsection{Effects of Reparameterization}
% Figure A for this is in fig-reparameterization.tex; placement is currently in experiments.tex
% Figure B for this is in fig-kernel.tex; placement is currently in experiments.tex

%%%%%%%%%%%%%% FIGURE PLACEMENT %%%%%%%%%%%%%%%%%%%%%%%%%%

%Figure showing kernel effect with k vs K
\begin{figure}
	\centering
	\includegraphics[width=1.0\linewidth]{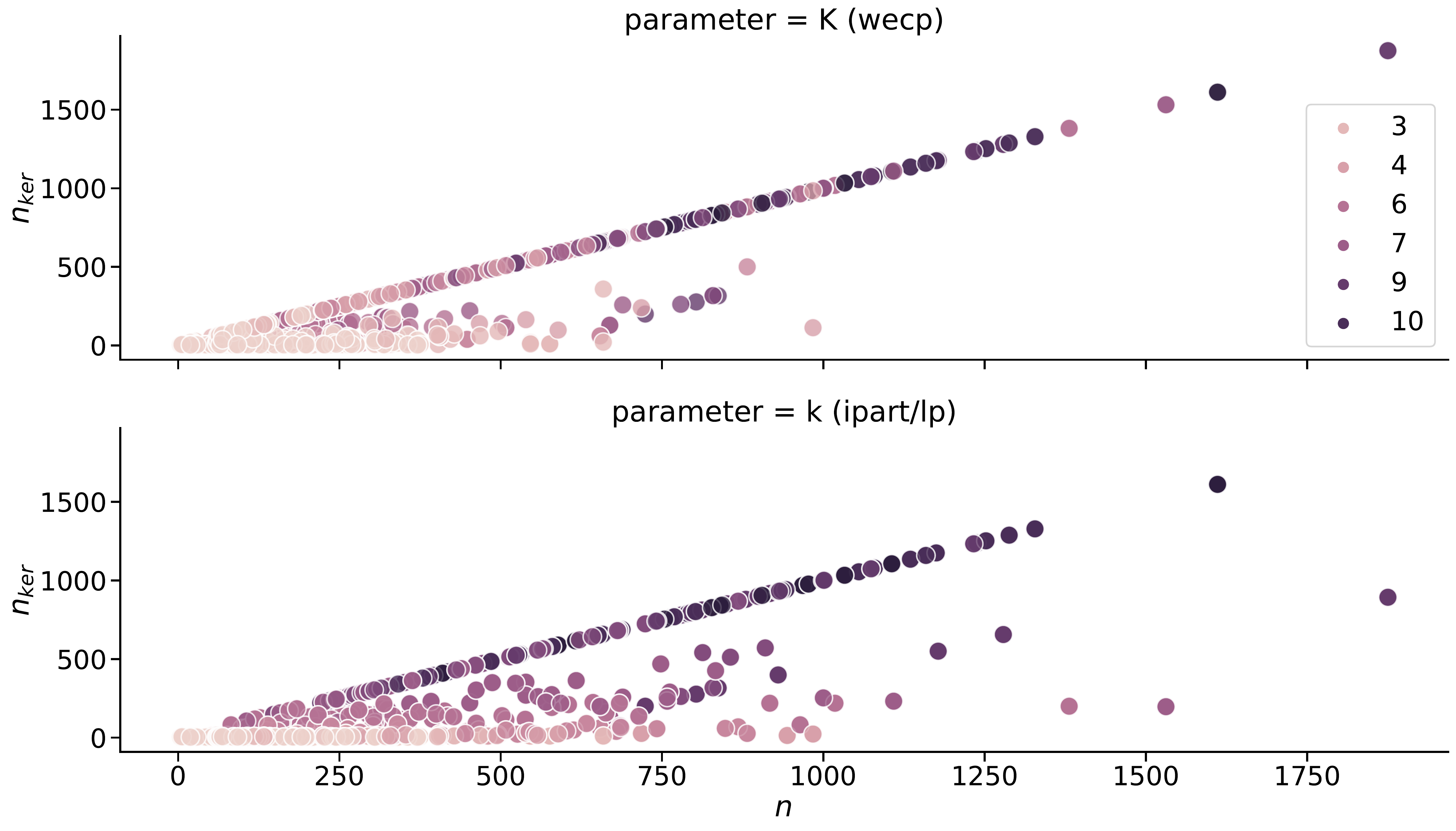}%
\caption{Instance size reduction due to kernelization (from $n$ to $n_{ker}$); points along the diagonal experienced no reduction from kernel rules. Top shows reduction using parameter $K$, bottom shows reduction when using parameter $k$.}%
\label{subfig:2c}%
\end{figure}

%%%%%%%%%%%%%%%%%%%%%%%%%%%%%%%%%%%%%%%%%%%%%%%%%%%%%%%%%%

%%%%%%%%%%%%%%FIGURE PLACEMENT %%%%%%%%%%%%%%%%%%%%%%%%%%

%Figure including all comparisons where we ran the Feldmann et al algorithm
\begin{figure}[h]
  \centering
  \includegraphics[width=1.0\linewidth]{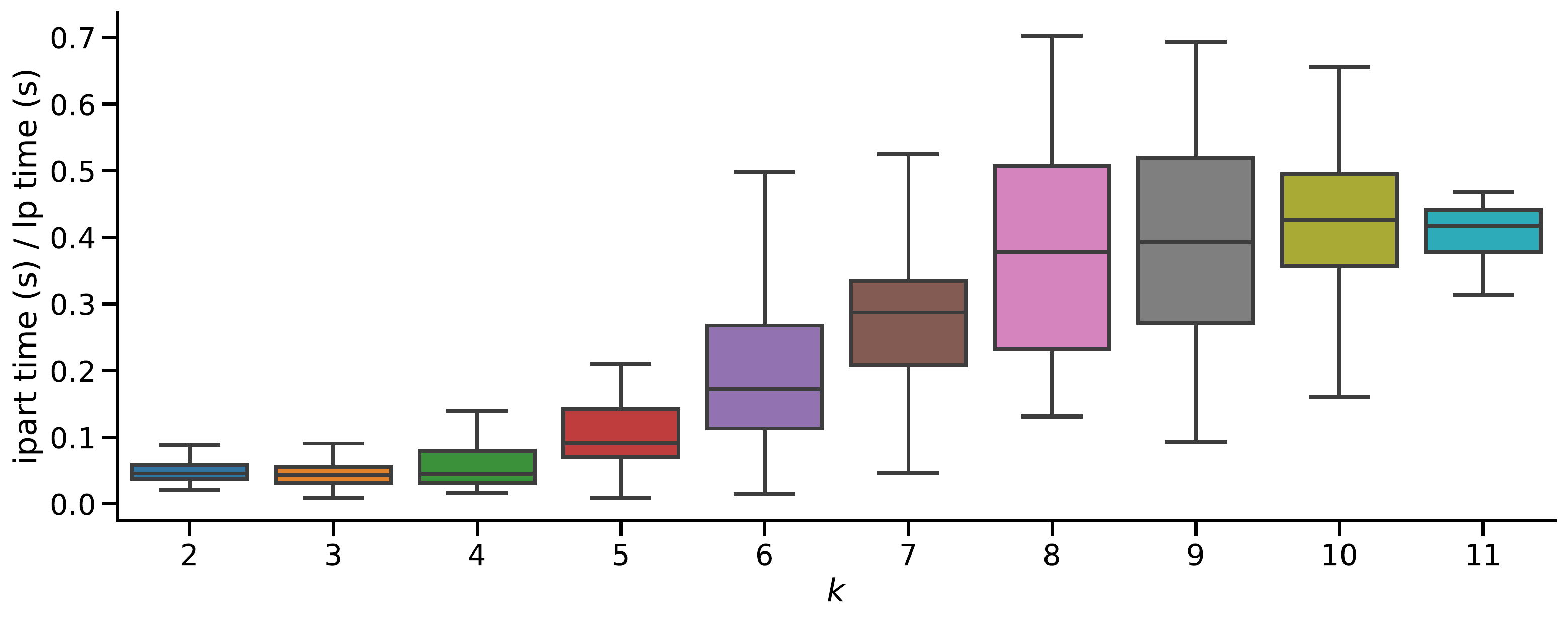}
  \caption{Relative runtimes of \ipalg and \lpalg on full corpus with $2 \leq k \leq 11$ and all weight scalings; times exclude kernel (which is shared). Outliers not shown.}
\label{fig:ip-v-lp}
\end{figure}

%%%%%%%%%%%%%%%%%%%%%%%%%%%%%%%%%%%%%%%%%%%%%%%%%%%%%%%%%%

To compare the effects on the runtime of reparameterizing from the sum of the clique weights $K$ to the number of distinct cliques $k$, we tested \feldalg, \ipalg, and \lpalg on all instances with $k \leq 11$ and small/medium weight scalings. As seen in 
Figure~\ref{fig:orig-v-new-2a}, and Figures~\ref{fig:orig-v-new-2b},
 ~\ref{fig:all-log-kernelized}, ~\ref{fig:orig-v-new-full} in Appendix~\ref{appendix:expresults}, 
both algorithms parameterized by $k$ are faster across the entire corpus when $k>=6$. It should be noted that the slower \ipalg and \lpalg runtimes for $k<6$ are partly due to the preprocessing often removing highly weighted cliques, effectively setting $k=K$. When $k=K$, it is slightly faster to run \feldalg due to not having to compute the clique weights. 
Figure~\ref{fig:orig-v-new-2a} shows the runtimes for $K\in[2,20]$, and Figure
~\ref{fig:orig-v-new-full} shows all $K \in [2,49]$. After $K=13$ \feldalg timedout for every instance, whereas \ipalg and \lpalg are still able to compute solutions in under an hour\looseness-1. 

Some performance increase may be attributed to the kernelization, as shown in Figure~\ref{subfig:2c}. Since one of the reduction rules relies on the parameter, the instance size after kernelization (denoted $n_{ker}$) is different between \feldalg (which uses $K$) shown in the top figure, and \ipalg/\lpalg (which use $k$) shown in the bottom figure. Figure~\ref{subfig:2c} shows that the kernel gives great reduction when parameterized by $k$ when $k$ is small and has less of an impact when $k$ is large. 
Figure~\ref{fig:all-log-kernelized} in Appendix~\ref{appendix:expresults} shows a comparison of the runtimes when instances are sorted by the size of the kernelized instance.

\subsection{Integer Partitioning vs LP}
%Figure showing difference based on parameter is in fig-IPvLP.tex; placement is currently in experiments.tex

From Figure~\ref{fig:orig-v-new-2a}, we observe that \ipalg runs slightly faster on average than \lpalg, which was unexpected. To further compare the two, we ran both algorithms on a larger corpus including all instances with $k \leq 11$ regardless of weight scaling. Figure~\ref{fig:ip-v-lp} shows the runtime ratio of the two algorithms\footnote{these times exclude the shared kernel to emphasize the difference in the two approaches}. We observe that despite a consistent advantage for \ipalg on small $k$, the methods' runtimes seem to converge as we approach $k = 9$ and we hypothesize that \lpalg will become the dominant approach for larger $k$ when testing with a longer timeout than one hour. To test whether the specific clique-weight assignment mattered, we evaluated runtimes on $6$ random weight assignment permutations for each instance. Table~\ref{table:weightperm} in 
Appendix~\ref{appendix:varyweight} shows that both algorithms are virtually unaffected by the assignments\looseness-1.

Finally, since \ipalg's complexity depends on the maximum weight $w$, we evaluated this effect by comparing performance across the small, medium, and large weight-scale variants of each instance. Figure~\ref{fig:maxweight-diff} in Appendix~\ref{appendix:expresults} shows that the relative increase in runtime between weight scalings is fairly consistent across algorithms, indicating minimal effect. However,
it is noteworthy that \ipalg experiences much higher variance\looseness-1.

\subsection{Ground Truth}
While these algorithms are guaranteed to find a decomposition using at most $k$ weighted cliques if one exists, a unique solution is not guaranteed. For the synthetic corpus, we verified that our output recovered the LVs/TFs selected by the generators, but we do not know whether this will generalize to real data (where weight distributions may be quite different) or much larger $k$.
Additionally, Appendix~\ref{appendix:guessk} analyzes the impact on incorrect $k$ input on the recovered solution.

\section{Conclusions \& Future Work}
This paper offers a new combinatorial framing of the problem of module identification in gene co-expression data, \WCDfull. Further, we present two new parameterized algorithms for the noise-free setting (\EWCD), removing the dependence of a prior approach on the magnitude of the clique weights. To address concerns of practicality, we implement both approaches and evaluate them on a corpus of biologically-inspired genetic association data. The empirical results show that both new approaches significantly outperform the \WECPfull algorithm of~\cite{feldmann2020}, and that worst-case asymptotic runtime bounds are not realized on typical inputs.

While these algorithms provide a nice first step towards a polynomial-time algorithm for module-identification, there remain many unaddressed challenges before use on real data. While our exponential dependence on the number of modules is to be expected from an FPT algorithm, in many real-world datasets there are hundreds of underlying functional groups. One way to overcome this limitation is to incorporate a hierarchical approach, which would provide an extra biologically meaningful outcome: the relationships between cliques could be mapped to representations such as the Gene Ontology, where descendents represent more specialized function.

We would like to extend our approach to the non-exact setting, targeting approximation schemes for the optimization variant of the problem. Understanding the correct penalty function for under- and over-estimating edges, and whether this should be amplified for edges below the threshold in the original coexpression data will be critical in informing a useful technique.\looseness=-1

\pagebreak 

\appendix
\section{Kernel for \AEWCD}\label{appendix:kernel}
\label{sec:kernel}
Here, we give the $4^k$-kernelization for \WBSDDW (and hence also for the equivalent \AEWCD), and prove its correctness and runtime.
Let $(A,k)$ be the input \WBSDDW instance and let $(G,w_e,w_v,k)$ be the corresponding \AEWCD instance.
For a vertex $u$ of $G$ we will use $A_u$ and $B_u$ to denote their corresponding rows in matrices $A$ and $B$ respectively. Similarly we use $A_{uv}$ for an element of the matrix $A$ corresponding to the pair of vertices $u$ and $v$.

First, we divide the vertices of input graph $G$ (correspondingly the rows of input matrix $A$) into \emph{blocks}, as follows.
We say that $i$ and $j$ are in the same block if $A_i\es A_j$.
Note that $\es$ is an equivalence relation over the rows of $A$, as proven by Feldmann et. al.~\cite[Lemma 7]{feldmann2020}.
Then, we apply the following two reduction rules exhaustively.
\begin{reduction}
	\label{rr:No}
	If there are more than $2^k$ blocks then output that the instance is a NO-instance.
\end{reduction}

\begin{reduction}
	\label{rr:block}
	If there is a block $D$ of size greater than $2^k$, then pick two distinct $i,j\in D$. We reduce to an instance $(A',k)$ as follows: $G':=G- \left(D\setminus\left\{ i \right\}\right)$, $A'_{ii}=A_{ij}$, and $A'_{uv}=A_{uv}$ for all $\left( u,v \right)\in \left(V(G')\times V(G')\right) \setminus \{\left( i,i \right)\}$.  
	Given a solution $(B',W')$ for $(A',k)$ we construct a solution for $(A,k)$ as $W=W'$, $B_u=B'_u$ for all $u\notin D$, and $B_u=B'_i$ for all $u\in D$.
\end{reduction}

Once the two rules are applied exhaustively, then the reduced instance has size at most $4^k$ because there are at most $2^k$ blocks by \cref{rr:No} 
and each block size is at most $2^k$ by \cref{rr:block}.
So, it only remains to prove that the two reduction rules are correct, and also to prove the runtime of the kernelization.
The following lemma gives the correctness of \cref{rr:No}.

\begin{lemma}
	\label{lem:correctness-rr1}
	If $(A,k)$ is a YES-instance of \WBSDDW, then there are at most $2^k$ blocks in $A$.
\end{lemma}
\begin{proof}
	Suppose there are more than $2^k$ blocks.
	Let $(B,W)$ be a solution.
	Since there are only $2^k$ distinct binary vectors, there
	exist $i$ and $j$ in different blocks such that $B_i=B_j$. 
	Then we have $A_i\es B_i^TWB_i=B_j^TWB_j\es A_j$. 
	This implies $A_i\es A_j$ (because if $x\es y\es z$ and $y$ does not contain any $\star$ then $x\es z$), and hence $i,j$ and $j$ are in the same block, a contradiction.
\end{proof}

Now, we prove the correctness of \cref{rr:block} in the following two lemmas.

\begin{lemma}
	\label{lem:correctness-rr1-yes}
	In \cref{rr:block}, 
	if the reduced instance $(A',k)$ has a solution $(B',W')$ then the solution $(B,W)$ constructed by \cref{rr:block} is indeed a solution to $(A,k)$.
\end{lemma}
\begin{proof}
	It is sufficient to prove that $B_u^TWB_v\es A_{uv}$ for all $u,v\in V(G)$.	
	First consider the case when $u,v\notin D$, the block picked by \cref{rr:block}.
	Then $B_u^TWB_v={B'_{u}}^{T}W{B'_{v}}\es A'_{uv}=A_{uv}$.
	Now, consider the case when $u\in D,v\notin D$.
	Then $B_u^TWB_v={B'_u}^TWB'_i\es A'_{ui}=A_{ui}$. 
	Finally, consider the case when $u\in D,v\in D$.
	We can assume $A_{uv}\neq \star$ as this case follows trivially.
	Then $B_u^TWB_v={B'_i}^TWB'_i\es A'_{ii}=A_{ij}= A_{uv}$. 
	Here, the last equality is because any two entries (that are not $\star$) in the same block of matrix $A$ are equal \cite[Lemma 7]{feldmann2020}.
\end{proof}

\begin{lemma}
	\label{lem:correctness-rr1-no}
	In \cref{rr:block}, 
	if $(A,k)$ is a YES-instance then
	the reduced instance $(A',k)$ is a YES-instance.
\end{lemma}
\begin{proof}
	Let $(B,W)$ be a solution of $(A,k)$.
	Since the block $D$ contains more than $2^k$ rows, there exist row indices $p$ and $q$ such that $B_p=B_q$.
	We define a solution $(B',W)$ for $(A',k)$ as $B'_u:=B_u$ for all $u\in V(G')\setminus\left\{ i \right\} $ and $B'_i:=B_p$.

	To prove that $(B',W)$ is indeed a valid solution for $(A',k)$,
	it is sufficient to show that ${B'_u}^TWB'_v\es A'_{uv}$ for all $u,v\in V(G')$.	
	First consider the case when $u,v\neq i$.
	Then ${B'_u}^TWB'_v={B_{u}}^{T}W{B_{v}}\es A_{uv}=A'_{uv}$.
	Now, consider the case when $u=i$,$v\neq i$.
	Then ${B'_i}^TWB_v={B_p}^TWB_v\es A_{pv}=A_{iv}=A'_{iv}$, 
	where the second-to-last equality followed as $p$ and $i$ are in the same block $D$.
	Finally, consider the case when $u=v=i$.
Then ${B'_i}^TWB'_i={B_p}^TWB_p=B_p^TWB_q= A_{pq}\es A_{ij}= A'_{ii}$. 
	Here, the $\es$ follows because any two entries (that are not $\star$) in the same block of matrix $A$ are equal \cite[Lemma 7]{feldmann2020}.
	Also, note that the third equality is an equality (and not only a `$\es$ equivalence') as $A_{pq}$ is not a diagonal entry.
\end{proof}

It is rather easy to see that the runtime of the kernelization is $\mathcal O(n^3)$.
The division into blocks can be easily realized in $\mathcal O(n^3)$ time.
The application of \cref{rr:No} then takes only $\mathcal O(1)$ time.
\cref{rr:block} is applied at most once to each block and all the applications together take only $\mathcal O(n^2)$ time.

\section{Algorithm Correctness}\label{appendix:algcorrect}
\subsection{\AlgLp}
\label{section:missingLP}
Here we give the missing proofs for the correctness of algorithm {\AlgLp}.
\begin{proof}[Proof of \cref{lem:YesReturn}]
	Each row of $B$ is either a pseudo-basis row that was filled in \cref{line:LpBasisFill}	of {\AlgLp} or it is a non-basis row that was filled in \cref{line:BFill} of {\FillNB}.
	Now consider two pseudo-basis rows $B_i$ and $B_j$.
	For them, we have $B_i^TWB_j= A_{ij}$ as $W$ was a solution to the LP that contained the constraint $B_i^TWB_j=A_{ij}$.
	Also, for a pseudo-bais row $B_i$, we have that $B_i^TB_i\es A_{ii}$ because if $A_{ii}\neq \star$, then we added the constraint $B_i^TWB_i=A_{ii}$ to the LP.
	Now consider a non-basis row $B_i$ and some other row $B_j$ that was filled before $B_i$.
	Note that $B_j$ could be a pseudo-basis row or a non-basis row.
	Since the algorithm filled $B_i$ in \cref{line:BFill} of {\FillNB},
	we know that $B_i$ is $(i,W)$-compatible with all the rows filled before.
	Thus $B_i^TWB_j=A_{ij}$.
	Moreover, by $(i,W)$-compatibility, we also have $B_i^TWB_i\es A_{ii}$.
	Hence, we have that $BWB^T\es A$.
\end{proof}

\subsection{\AlgIp}\label{appendix:IP-correct}
 As noted in the main text, the majority of the proof of {\AlgIp}'s correctness follows the proof of {\AlgLp} in \cref{section:Lpcorrectness}, so here we only point out the differences.

Since the main difference with the LP algorithm is the {\InferCWIp} routine, we prove the correctness of it in the following lemma.
The lemma follows directly from the construction of {\InferCWIp} and {\UpdWsIp} algorithms.
\begin{lemma}
	\label{lem:updatecorrectness}
	Consider the function call {\InferCWIp}($A$, \allowbreak$\basis$, $\textbf{W}$, $i$) in \cref{line:IpInferCall} of {\AlgIp}.
	Let $\basis'$ be the matrix $\basis$ before the insertion of current row $\basis_i$, i.e. $\basis'_i$ is a null row.
	Suppose $\textbf{W}$ contains all the compatible $W$ with $\basis'$ then the list $\textbf{S}$ returned
	contains all the compatible $W$ with $\basis$.\looseness-1
\end{lemma}

Once we have the above lemma, the correctness of the algorithm follows more or less the same proof as that of the LP algorithm.
We state the following two key lemmas that are counterparts of
\cref{lem:YesReturn} and \cref{lem:iextend}.
\begin{lemma}
	If {\AlgIp} returns through line~\ref{line:IpYesReturn}, then the matrices $B$ and $W[0]$ output satisfy that $A\es BW[0]B^T$.
\label{lem:IpYesReturn}
\end{lemma}

\begin{lemma}
	If \FillNB$(A,\basis=B^*_I,W[0])$ called on \cref{line:IpFillNBCall} of {\AlgIp} returns $i\le n$, then $B^*_i$ $i$-extends $B^*_I$.
	\label{lem:iextendIp}
\end{lemma}
Both the lemmas follow the same proof as that of their counterparts.
To derive \cref{lem:iextendIp} from the proof of \cref{lem:iextend}, it is sufficient to observe that the statement, \emph{all solutions of the LP still remain solutions},
can be interpreted as all compatible $W$'s still remain compatible.
This is the reason why we need only to check with $W[0]$ in {\FillNB} and not with all matrices in the list $\textbf{W}$.
In fact, this is what we do even in the LP; the LP has many possible solutions but we check with only one solution.
The difference is that the many solutions are implicitly captured by the LP constraint system, whereas \AlgIp explicitly maintains all compatible combinations.

\section{Runtime Analysis}\label{appendix:algruntime}
Here, we estimate the runtimes of {\AlgLp} and {\AlgIp}.
Note that the runtimes we give are after kernelization, i.e. the input to these two algorithms are assumed to be a kernel according to Theorem~\ref{thm:kernel}.
The kernelization incurs an additional runtime additive factor of $\mathcal O(n^3)$.
\subsection{\AlgLp}\label{section:Lpruntime}

\begin{lemma} \label{thm:Lpruntime} {\AlgLp} (Algorithm~\ref{alg:Lpmain})
	runs in time
	$\mathcal O(4^{k^2}k^2(32^k+k^3 L))$,
 where $L$ is the number of bits required for input representation.
\end{lemma}

\begin{proof}
The \for loop in \cref{line:LpFor} has at most $2^{2k^2}$ iterations.
The \while loop in \cref{line:LpWhile} has at most $2k$ iterations.
The only steps that take more than unit time in the \while loop are the calls to {\InferCWLp} and {\FillNB}.
{\InferCWLp} solves an LP with $k$ variables and at most $4k^2$ constraints.% and the largest entry being $\|A\|_{\infty}$.
This can be solved in at least $\mathcal O(k^4 L)$ time by using standard algorithms~\cite{LpVaidya89}.
It only remains to estimate the runtime of {\FillNB}.
The \while loop in \cref{line:fillnbwhile} of {\FillNB} has at most $n$ iterations and the \for loop in \cref{line:fillnbfor} has at most $2^k$ iterations.
The only non-trivial step in the \for loop is the call to \Comp.
The \for loop in \cref{line:compfor} of {\Comp} has at most $n$ iterations and \cref{line:compmulti} takes at most $k$ operations.
Thus the time taken for {\Comp} is at most $\mathcal O(nk)$ and the time taken for {\FillNB} is at most $\mathcal O (n^2k2^k)$.
Hence, the time taken for {\AlgLp} is in $\mathcal O(2^{2k^2}2k(k^4 L+n^2k2^k))$.
The claimed run-time in the theorem follows by putting $n\le 4^k$ due to the kernel.
\end{proof}

\subsection{\AlgIp}\label{section:Ipruntime}

\begin{lemma} \label{thm:Ipruntime} {\AlgIp} (Algorithm~\ref{alg:Ipmain})
	runs in
	$\mathcal O(4^{k^2} 32^k w^kk)$, 
	where $w$ is the maximum weight value in $A$. 
\end{lemma}

\begin{proof}
	The \texttt{for} loop in \cref{line:IpFor} has at most $2^{2k^2}$ iterations.
	Let $y$ be the number of distinct partially filled weight matrices. 
	We have $y\le (w+2)^k$ since each such matrix is defined by the $k$ entires along its diagonal, each of which is either null or an integer from $0$ to $w$.

	Now, we show that for a fixed iteration of \texttt{for} loop in \cref{line:IpFor} of {\AlgIp}, 
	\begin{enumerate}
		\item
	each line of {\AlgIp} in the \texttt{for} loop is executed at most $y$ times,
		\item
	each line in {\InferCWIp} is executed at most $y$ times (across all calls to the function from the fixed iteration of for loop), and 
		\item
			the \texttt{for} loop in \cref{line:ForLoopParts} of 	
			{\UpdWsIp} is executed at most $y$ times (across all calls to {\UpdWsIp} from all calls of {\InferCWIp} from the fixed iteration of for loop in {\AlgIp}).
	\end{enumerate}
	
	For (1) observe that the list $\textbf{W}$ in {\AlgIp} always gets new matrices whenever it is modified.
	At any point the elements of the list are disjoint from the past entries of the list (during the fixed iteration of for loop).
	For (2) observe that between the two consecutive executions of a line, \textbf{temp} would have seen a new matrix that was not in it before.
	For (3) observe that every time a new matrix is pushed to $\textbf{V}$, it is a new matrix that it did not have before.

	From \cref{section:Lpruntime}, we know that {\FillNB} runs in $\mathcal O(n^2k2^k)$ time. 
	All the other non-loop lines in {\AlgIp}, {\InferCWIp} and {\UpdWsIp} can be done in at most $\mathcal O(k^2)$ time.
	The for loop in \cref{line:InnerForLoopUpdateWs} takes only $\mathcal O(k)$ time.

Therefore, the total running time of {\InferCWIp} is 
\begin{align*}
&\mathcal O(2^{2k^2}\cdot w^k\cdot (n^2k2^k+k^2))\\
&=\mathcal O(4^{k^2} 2^kw^kn^2k)\\
&=\mathcal O(4^{k^2} 32^kw^kk)
\end{align*}

where the last equality follows by using that $n\le 4^k$ after kernelization.
\end{proof}

%Figure including comparisons where we ran the Feldmann et al algorithm
\begin{figure*}
	\centering
	\includegraphics[width=1.0\linewidth,height=4.3cm]{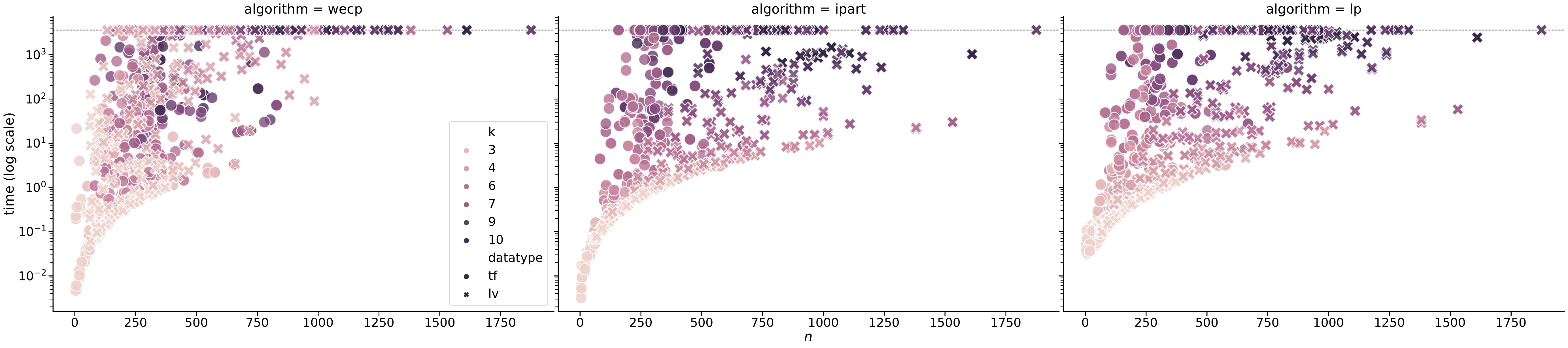}%
\caption{Log-scale runtimes (kernel $+$ decomposition) of \feldalg, \ipalg, and \lpalg on corpus of all TF (circle) and LV (star) instances with $2 \leq k \leq 11$ and small/medium weight scalings, sorted by instance size (prior to running the kernel).}%
\label{fig:orig-v-new-2b}%
\end{figure*}

\begin{figure*}[h]	
  \includegraphics[width=1.0\linewidth,height=4.3cm]{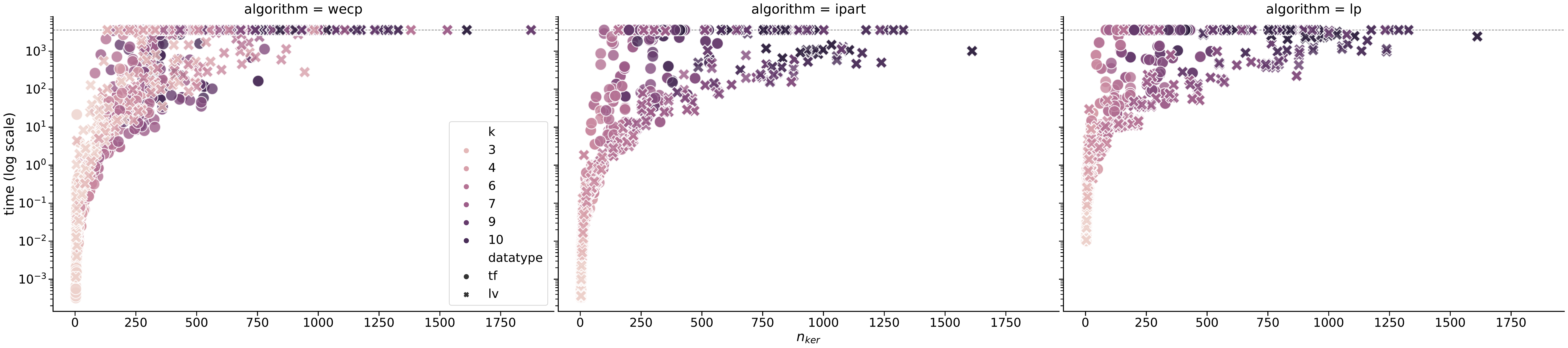}
  \caption{Log-scale runtimes (decomposition only) of \feldalg, \ipalg, and \lpalg on corpus of all TF (circle) and LV (star) instances with $2 \leq k \leq 11$ and small/medium weight scalings, sorted by reduced instance size (after running the kernel).}
\label{fig:all-log-kernelized}
\end{figure*}

\begin{figure}[h]
  \includegraphics[width=\linewidth,height=6.5cm]{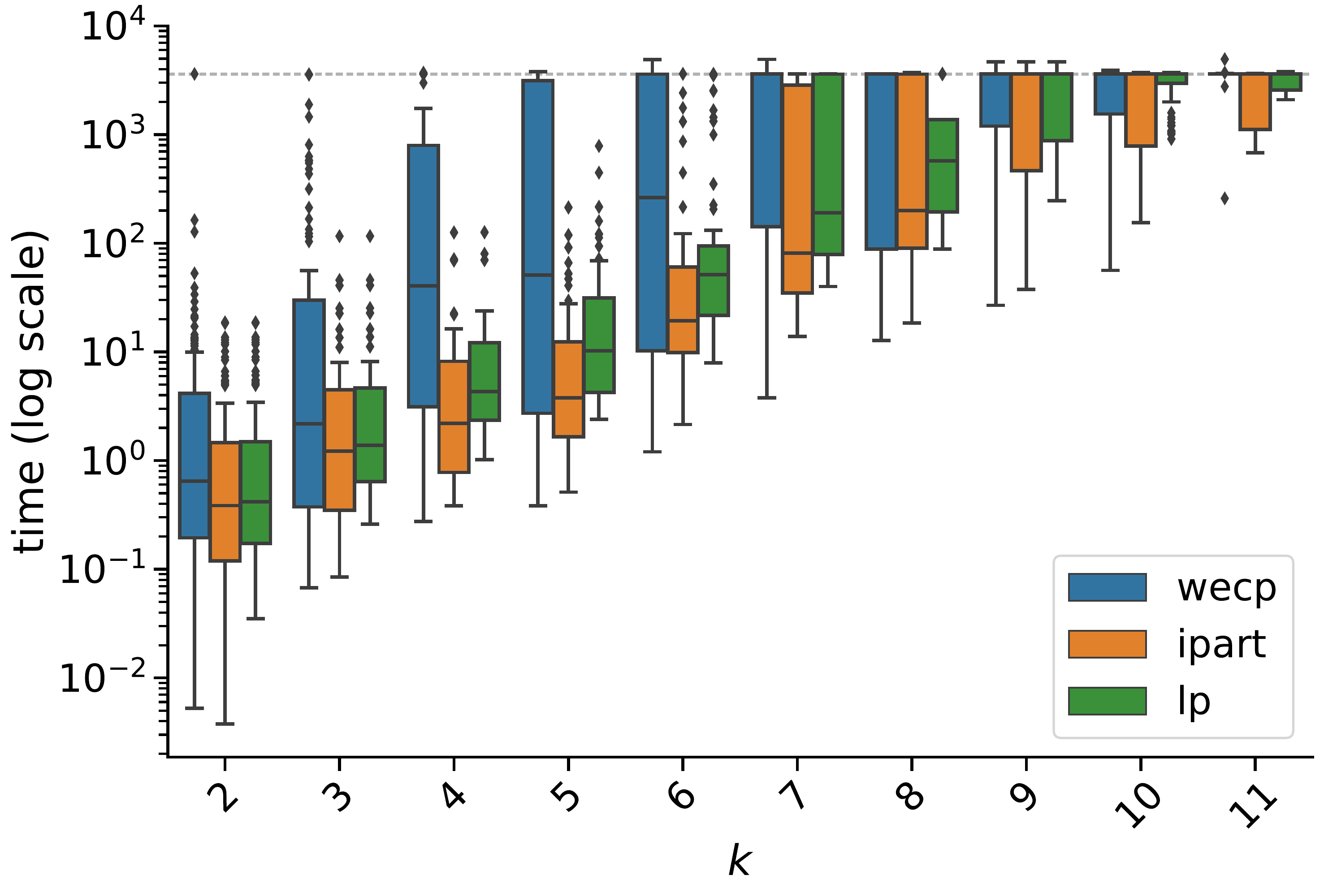}
  \caption{Log-scale distribution of total runtimes (preprocessing + kernel + decomposition) binned by $k$ on all TF and LV instances with $2 \leq k \leq 11$ for small/medium weight scalings.}
\label{fig:all-box-k}
\end{figure}

\begin{figure*}[h]
  \includegraphics[width=\linewidth,height=5.5cm]{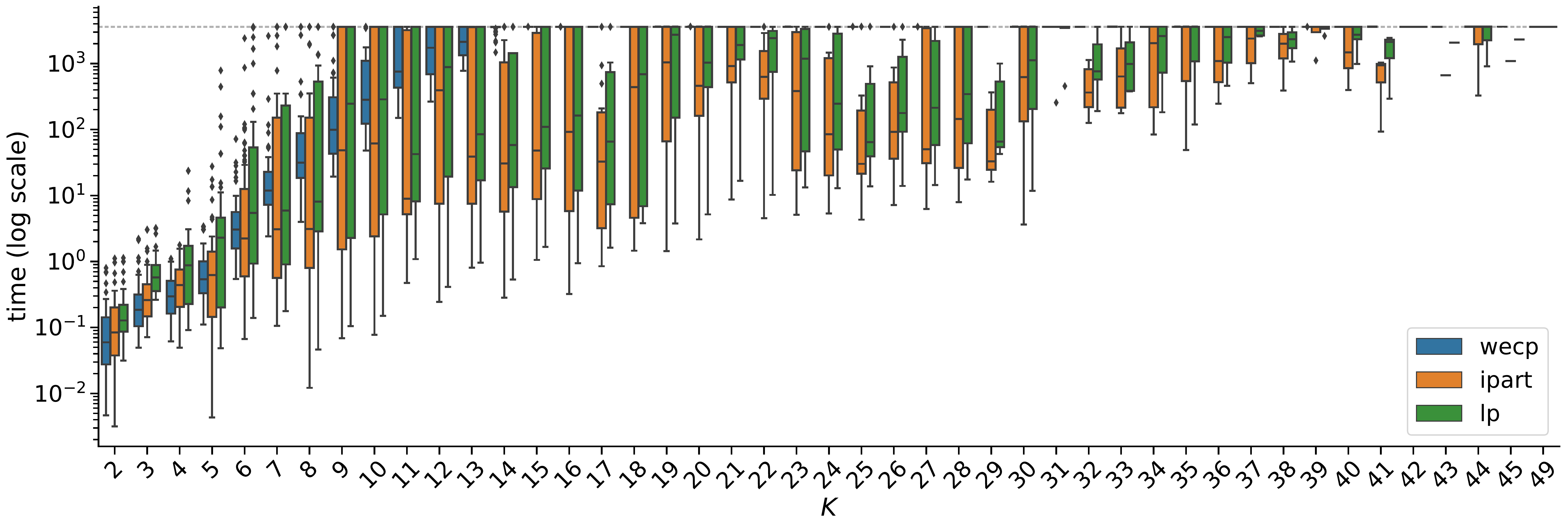}
  \caption{Log-scale plot showing distribution of total algorithm runtimes (kernel + decomposition) when binned by $K$ for all $K$ values.}
\label{fig:orig-v-new-full}
\end{figure*}

\begin{figure*}[h]
  \includegraphics[width=\linewidth,height=4.5cm]{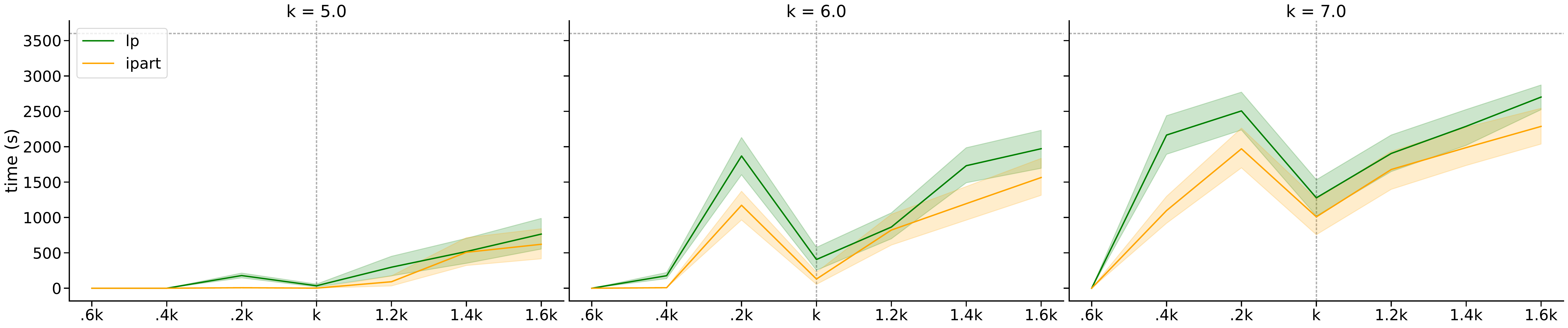}
  \caption{Average runtimes (kernel + decomposition) between all instances with ground truth $k=5,6,7$ when inputting $k$ values in the range $[0.6k, 0.4k, 0.2k, k, 1.2k, 1.4k, 1.6k]$. Vertical dashed line indicates average time for ground-truth $k$ as input.}
\label{fig:guessk}
\end{figure*}

\section{Synthetic Data Specifications}\label{appendix:synthdata}
\begin{table}[h]
\centering
\begin{tabular}{l|rrr|rrr}
\toprule
    &        \multicolumn{3}{c|}{LV}   &   \multicolumn{3}{c}{TF}      \\
    $k$ &  \# & \multicolumn{1}{c}{$n$} &  \multicolumn{1}{c}{$m$} &  \# & \multicolumn{1}{c}{$n$} & \multicolumn{1}{c}{$m$} \\
\midrule
2              &   60 &   129.3 &    6325.2 &   64 &   36.4 &   1221.7 \\
3              &   60 &   229.3 &   17571.3 &   67 &   67.7 &   2315.7 \\
4              &   60 &   352.6 &   32644.5 &   65 &   83.0 &   2750.1 \\
5              &   60 &   387.8 &   41867.6 &   65 &  111.8 &   5434.5 \\
6              &   60 &   445.0 &   42498.8 &   69 &  175.0 &  16745.6 \\
7              &   60 &   605.5 &   92604.6 &   65 &  113.0 &   3212.9 \\
8              &   60 &   661.6 &   89747.0 &   63 &  222.9 &  15585.9 \\
9              &   60 &   713.7 &  110952.1 &   72 &  233.8 &  19834.6 \\
10             &   60 &   737.2 &   74706.4 &   67 &  269.9 &  17520.0 \\
11             &   60 &   823.1 &   77260.8 &   63 &  222.7 &  13646.4 \\
12             &   45 &   876.8 &  112389.7 &   70 &  272.3 &  20116.5 \\
13             &   42 &   872.9 &   69691.5 &   63 &  361.8 &  35362.8 \\
14             &   45 &   859.9 &   69564.9 &   55 &  286.7 &  10791.9 \\
15             &   27 &   954.6 &   66220.1 &   59 &  386.4 &  38712.8 \\
16             &   12 &   848.2 &   45401.5 &   65 &  319.9 &  18656.3 \\
17             &    - &     - &       - &       43 &  361.0 &  16164.7 \\
18             &    6 &   897.5 &   44828.5 &   37 &  354.6 &  24494.3 \\
19             &   21 &  1042.7 &   57171.1 &   24 &  345.0 &  13114.5 \\
20             &    3 &  1029.0 &   41588.0 &   40 &  340.4 &  13205.0 \\
\bottomrule
\end{tabular}
\caption{Average instance sizes (number of nodes $n$ and edges $m$)  across $k$ values for TF and LV datasets.
The number of graphs for a given $k$ value may be smaller than 60 if some instances were eliminated for not having $2-11$ cliques after preprocessing.}
\label{table:datastats}
\end{table}

Here we detail the parameter settings and methodology for the synthetic corpus generation. Each instance is generated by specifying a random seed, a desired number of cliques $k$, one of three weight scaling factors, and an underlying dataset (TF or LV). For each combination of parameters selected, we used 20 random seeds. We used all $k$ values in $[2,20]$, resulting in an initial corpus of $2280$ graphs. We only ran experiments on those instances which had $k$ values in $[2,11]$ after pre-processing (see Appendix~\ref{appendix:preprocessing}), resulting in a final corpus of $1917$ networks. Table~\ref{table:datastats} summarizes the average number of nodes and edges for the generated corpus.

\begin{table}[h]
\centering
\begin{tabular}{l | rrr | rrr}
\toprule
% node overlap   |   clique overlap
{} & \multicolumn{3}{c}{Node Overlap} & \multicolumn{3}{|c}{Clique Overlap} \\
{} &  min. &  avg. &  max. &  min. &  avg. &  max. \\
\midrule
0-20\%   &          1911 &          1279 &           722 &            1785 &             917 &             311 \\
21-40\%  &             6 &           568 &           397 &             108 &             787 &             405 \\
41-60\%  &             0 &            70 &           266 &              21 &             201 &             672 \\
61-80\%  &             0 &             0 &           246 &               3 &              12 &             410 \\
81-100\% &             0 &             0 &           286 &               0 &               0 &             119 \\
\bottomrule
\end{tabular}
\caption{Summary of overlap between cliques across the corpus. At left, we report on the number of nodes shared between cliques in each graph (as a percent of $n$). The first column (min) reports the number of graphs where every clique had at least the given amount of overlap, the second (avg) reports based on the average, and the third (max) gives the number of graphs where the clique with the most overlap fell into the given range. At right, we use the same min, avg, max criterion, but instead measure overlap by the percentage of other cliques sharing at least one vertex. }
\label{tab:clique-overlap}
\end{table}
Table~\ref{tab:clique-overlap} summarizes statistics about how the ground truth cliques overlapped across the entire corpus. For example, about six percent of the graphs ($119$ of the $1917$) contained at least one clique which overlapped almost all ($81-100\%$) of the remaining cliques, and over ten percent ($201$) have their average clique overlapping over $40\%$ of all the other cliques. Alternatively, if you measure entanglement of cliques (modules) by the percentage of their nodes (genes) that are shared with at least one other clique, we can see that fifteen percent ($286$) have some clique which shares more than $80$ percent of its nodes with another clique, and just less than four percent ($70$ graphs) have an average overlap greater than $40\%$ for all their cliques.

\subsection{Clique Weight Scaling:}\label{appendix:dataSML}
In both corpora, we use three different scale factors (which we refer to as small, medium, and large) to control the maximum edge weight in the resulting networks. In the TF data, this takes the form of three different maximum edge weight values for our heavy-tailed weight generator: $1$, $4$, and $16$. In the LV data, we scale the average gene-LV association scores by $1$, $2$, and $4$. When this results in a non-integral weight; we take the ceiling.

\subsection{TF Weight Generation}\label{appendix:dataTF}
As noted in Section~\ref{section:tfdata}, the transcription factor dataset provides no inherent strength of association for each TF. Given the belief that real data follows a heavy-tailed distribution, we generate random integer weights that mimic this (to the extent possible, given the extremely small number of cliques to be assigned weights) as follows.

We take as input a desired maximum weight $\Delta$, and define three intervals $L = [1, \ell\Delta]$, $M = [m_l\Delta, m_r\Delta]$, $H = [h\Delta, \Delta]$. We set $\ell = .14$, $m_1 = .20$, $m_2 = .28$, and $h = .90$, ensuring the ranges are well-separated. To create a heavy-tail, we assign different probabilities to a weight being drawn from each range, $p_L = .75$, $p_M = .15$, and $p_R = .1$. Once an interval is selected, the weight is chosen uniformly at random among integers in its range. Each clique weight is generated from this process independently at random, with the caveat that we ensure that some clique receives weight $\Delta$ (to avoid instances with no high-valued clique).

\subsection{LV Membership Thresholding}\label{appendix:dataLV}
The LV data includes an association score for each gene-LV pair; in order to identify strongly-associated genes to be included in the clique generated from a given LV, we use a uniform threshold. In order to determine an appropriate threshold for maintaining some variability of clique-size without including large numbers of spurious associations, we computed an elbow plot showing the number of genes per latent variable. This resulted in a threshold of $0.6$, which was used for all instances.

\section{Supplemental Experimental Results}\label{appendix:expresults}
In this section, we provide additional experimental results. Note that in all data and analysis (in the appendix and main paper), $k$ is referring to the parameter value of the instances \textit{after} preprocessing.

\subsection{Additional Reparametrization Effects}\label{appendix:additional-reparam}
Figure~\ref{fig:orig-v-new-2b} shows the combined runtime of the kernel and decomposition algorithms on each instance sorted by the instance size before kernelization and colored by $k$ value. \ipalg (middle) and \lpalg (right) roughly solve instances with the same $k$ value (regardless of $n$) in a similar amount of time. This is shown by the light to dark gradient from the bottom to the top of the figures. Whereas \feldalg (left) has no discernible gradient, meaning there is no connection between $k$ and the running time of this algorithm. This is as expected since the input parameter to \feldalg is $K$ and not $k$.

Figure~\ref{fig:all-log-kernelized} is similar to Figure~\ref{fig:orig-v-new-2b}, but instead shows the running time of only the decomposition algorithms sorted by the instance size after kernelization and colored by $k$ value. When compared to Figure~\ref{fig:orig-v-new-2b}, these figures show that the kernel algorithm dramatically reduces the size of instances with small $k$ and has a negligible effect on instances with large $k$ values. The light to dark gradient is again apparent in the \ipalg (middle) and \lpalg (right) figures, meaning the decomposition runtimes are dependent mainly on $k$. There is a slight increase in running time as $n_{ker}$ increases as well. While the running time increases as $n_{ker}$ increases for \feldalg (left), again, the colors are sporadic. 
\feldalg timed out on $337$ instances, whereas \lpalg timed out on $204$ instances and \ipalg timed out on $171$ instances (out of 986 total instances) indicated by the points running across the top lines in both figures.

Figure~\ref{fig:all-box-k} shows the total runtime of preprocessing, kernelization, and decomposition binned by $k$. This figure shows that the median runtime of \feldalg across all $k$ values is larger than both \ipalg and \lpalg and has more variability. The median runtime of \ipalg is also less than \lpalg. Figure~\ref{fig:orig-v-new-full} shows the runtime of the kernel and decomposition algorithms binned over all $K$ values. \ipalg and \lpalg can compute solutions up to $K=44$, whereas \feldalg consistently times out once $K \geq 13$.

\begin{figure}[h]
  \includegraphics[width=\linewidth,height=6.5cm]{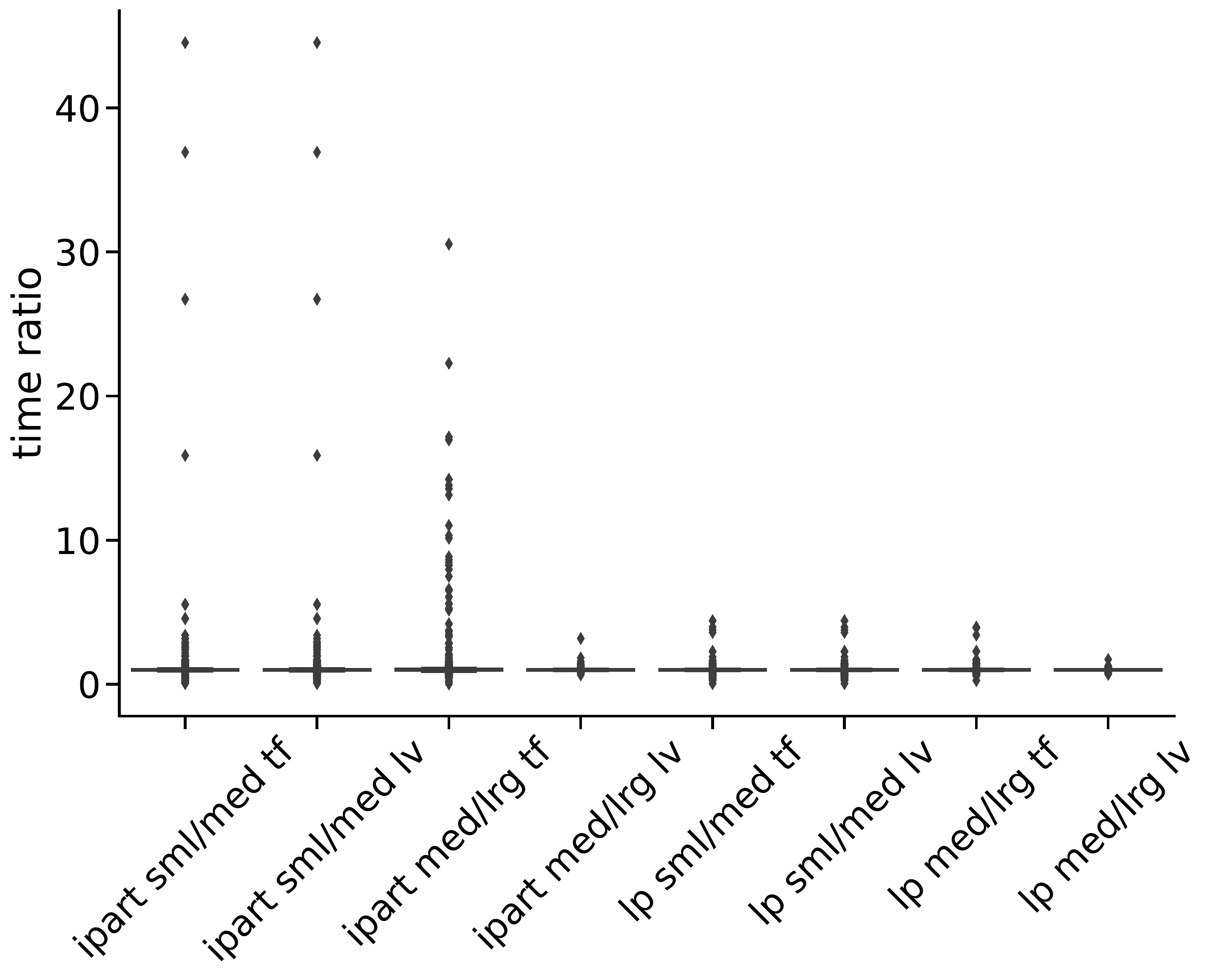}
  \caption{Runtime ratio of \ipalg and \lpalg when run on the same underlying instance with increased weight scaling factor: small vs medium and medium vs large.}
\label{fig:maxweight-diff}
\end{figure}

\begin{table}[h]
\centering
\begin{tabular}{l | cc}
\toprule
$k$ & \ipalg &  \lpalg \\
\midrule
2                  &  126.97  &  124.00 \\
3                  &  122.15  &  119.15 \\
4                  &  129.10  &  126.09 \\
5                  &  132.28  &  129.23 \\
6                  &  134.00  &  131.01 \\
\bottomrule
\end{tabular}
\caption{Average peak memory usage in megabytes between the \ipalg and \lpalg algorithms for $k$ between $[2-6]$. Data are combined for medium and large weight scales for both TF and LV datasets.}
\label{table:memoryusage}
\end{table}

Table~\ref{table:memoryusage} shows the average peak memory usage for graphs with medium and large weight scales for the entire corpus (combining both TF and LV graphs). We observe that the \ipalg algorithm's average memory usage is slightly larger than the \lpalg algorithm's, but the difference is not substantial. This contrasts with the \ipalg algorithm's large theoretical bound on the number of stored weight matrices. All $k$ values for both algorithms have a standard deviation around $25$. Peak memory usage was tracked for each run of the algorithms using the python \texttt{resource} module.

\subsection{Hardware}\label{appendix:hardware}
All experiments used identical hardware; each machine runs Arch Linux version 
$3.10.0-957.27.2.\text{el}7.\text{x}86\_64$, have $40$ Intel(R) Xeon(R) Gold 6230 CPUs (2.10GHz), and have $192$GB of memory. All code is written in Python 3.

\subsection{Varying Clique Weights}\label{appendix:varyweight}
\begin{table}[h]
\centering
\begin{tabular}{l | c | c}
\toprule
 $k$ & \ipalg & \lpalg      \\
\midrule
5 &                0.001 (0.004) &                0.000 (0.001) \\
6 &                0.019 (0.060) &                0.016 (0.066) \\
7 &                0.004 (0.012) &                0.071 (0.189) \\
8 &                0.000 (0.000) &                0.275 (0.287) \\
\bottomrule
\end{tabular}
\caption{Average difference (and standard deviation) between the maximum and minimum normalized runtimes when clique weight assignments were permuted six times per TF graph. Data are grouped by $k$. The runtime differences when $k \in [2,4]$ were all too small to be measured.}
\label{table:weightperm}
\end{table}

To test if clique weight assignments affect the runtime of the \ipalg and \lpalg algorithms, we randomly permuted the assignment of the same set of clique weights to the cliques of each TF graph six times. We ran both the \ipalg and \lpalg algorithms for each weight assignment and recorded the runtimes. To compare runtimes for each $k$ value, we first normalized all runtimes using min-max normalization. Using these normalized runtimes, we computed the difference between the maximum for a particular weight assignment and the minimum for each graph, then averaged across all graphs with the same $k$ value. Table~\ref{table:weightperm} shows these results. Since the differences are quite small, we conclude that specific clique weight assignments have little effect on the runtime of both the \ipalg and \lpalg algorithms.

\subsection{Performance When $k$ is Unknown}\label{appendix:guessk}
In practice, the ground truth number of distinct cliques (modules) in real-world graphs is unknown. We tested the effect of incorrect input of this parameter value on the runtime and solution quality.
Figure \ref{fig:guessk} shows the runtime ratio (including the kernel time) across the same instances when inputing $k$ not equal to the ground truth value. Inputting incorrect $k$ values for both \ipalg and \lpalg results in an large increase in runtime. When $k=0.6k$ (i.e., a much smaller value than the true $k$), the kernel outputs a No answer, thus the runtime of the decomposition algorithms is always zero. Interestingly, for all instances with
$k \in [0.6k, 0.4k, 0.2k]$, no solution was recovered, and for
$k \in [k, 1.2k, 1.4k, 1.6k]$, all ground truth solutions were recovered, even when the input $k$ value is larger than the ground truth.

\section{NP-hardness of \EWCD}\label{appendix:nphardness}
\label{app:NPhard}
We use the same construction as given by \cite{ma1988complexity}.
Using this, we will show that \EWCD is NP-hard even when restricted to $K_4$-free graphs and unit edge-weights.
They use a reduction from the NP-hard \probsty{Exact 3-Cover} (E3C).
In this problem,
we are given a universe $U$ of $3q$ elements and a collection $\Sets=\left\{ S_1,S_2,\dots,S_m \right\}$ of $m$ $3$-ary subsets of $U$.
The decision problem is whether there exist $q$ sets that cover all the elements.
Note that if there is such a covering then each element is covered exactly once in the solution.

\begin{figure}
	\centering
	\includegraphics[width=.8\linewidth]{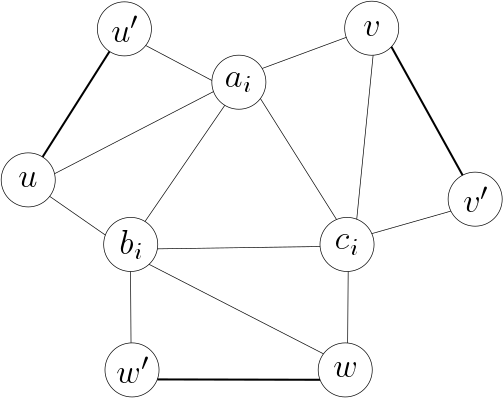}%
	\caption{The gadget for set $S_i=\{u,v,w\}$}%
\label{fig:NPhard}%
\end{figure}

Given an instance of E3C, we construct an instance of \EWCD on a $K_4$-free graph $G$ as follows.
For each element $u\in U$, we will have an edge $uu'$ in $G$.
We call these edges as the \emph{element-edges}.
For each set $S_i=\left\{ u,v,w \right\}$ in $\Sets$,
we will have three vertices $a_i,b_i,c_i$ that form a triangle.
We connect this triangle to the edges $uu'$, $vv'$ and $ww'$ as shown in \cref{fig:NPhard}.
All edges have weight $1$.
We set the budget $k$ for \EWCD to be $6m+q$.

First we show that if the E3C instance has a solution then so does the \EWCD instance.
Without loss of generality assume that $S_1,S_2,\dots S_q$ is a solution to E3C.
Then for each $1\le i\le q$,
we take the $7$ cliques $\{u,u',a_i\}$, $\{v,v',c_i\}$, $\{w,w',b_i\}$, $\{a_i,b_i,c_i\}$, $\{u,b_i\}$, $\{v,a_i\}$, and $\left\{ w,c_i \right\}$ into the solution.
For each $q+1\le i\le m$,
we take the $6$ cliques $\{u,a_i,b_i\}$, $\{v,a_i,c_i\}$, $\{w,c_i,b_i\}$, $\{u',a_i\}$, $\{v',c_i\}$, and $\{w',b_i\}$ into the solution.
We give each clique a weight of $1$.
It is easy to check that this gives a valid solution for \EWCD using exactly $k = 6m+q$ cliques.

Now, we show that if the \EWCD instance has a solution then so does the E3C instance.
Let $\C$ be the set of cliques in the solution to \EWCD.
Note that the cliques' weights could be fractional.
However, each edge has to be present in at least one of the cliques in $\C$.
Let $E_i$ denote the set of edges in the gadget corresponding to set $S_i$ that are exclusively in the gadget (the 12 thin edges in \cref{fig:NPhard}), that is, for $S_i=\{u,v,w\}$, the set $E_i=\left\{  a_ib_i, a_iu, ub_i, a_ic_i, a_iv, c_iv, b_ic_i,b_iw, wc_i,a_iu', c_iv',b_iw'\right\}$.
Let $\C_i$ be the set of all cliques in $\C$ which contain at least one edge from $E_i$.
It is easy to see that at least $6$ distinct cliques are required to cover the edge set $E_i$.
Hence, $|\C_i|\ge 6$.
Also, it is easy to see that $\C_i\cap \C_j=\varnothing$ for distinct $i,j$.
\begin{lemma}
	\label{lem:cliq7}
	If a clique $C\in \C_i$ contains an element-edge then $|\C_i|\ge 7$.
\end{lemma}
\begin{proof}
	Let $S_i=\left\{ u,v,w \right\}$.
	Without loss of generality, let the element-edge contained in $C$ be $ww'$.
	Then $C=\left\{ w,w',b_i \right\}$.
	Then $\C_i$ contains the $K_2$ $\left\{ c_i,w \right\}$.
	This is because the only other clique that could cover the edge $c_iw$ is $\left\{b_i,c_i,w  \right\}$; but this clique can have a weight strictly less than $1$ as otherwise the total weight of the cliques containing edge $b_iw$ exceeds $1$.
	Further, the edges in $E_i\setminus \left\{ b_iw',b_iw,wc_i \right\}$ require at least $5$ cliques to cover, proving
$|\C_i|\ge 7$.
\end{proof}

Since $k$ is only $6m+q$, the above lemma implies that there are $q$ indices $i\in [m]$ such that $\C_i$ covers $3$ element-edges.
Taking the sets $S_i$ for these $q$ indices gives the required solution for E3C.

\section{Preprocessing Specification}\label{appendix:preprocessing}
\begin{table}[h]
\centering
\begin{tabular}{lrrrr}
\toprule
$k$ &  \multicolumn{1}{c}{$n$}   &      \multicolumn{1}{c}{$m$}  &      \multicolumn{1}{c}{$k$}  & \multicolumn{1}{c}{\makecell{ cliques \\ removed }} \\
\midrule
2              &  92.74\% &  92.74\% &  92.74\% &      1.85 \\
3              &  67.45\% &  65.09\% &  72.97\% &      2.19 \\
4              &  65.78\% &  62.98\% &  73.60\% &      2.94 \\
5              &  51.08\% &  46.98\% &  66.56\% &      3.33 \\
6              &  36.53\% &  31.41\% &  53.75\% &      3.22 \\
7              &  44.59\% &  36.08\% &  62.29\% &      4.36 \\
8              &  36.30\% &  30.74\% &  52.95\% &      4.24 \\
9              &  36.43\% &  31.47\% &  52.95\% &      4.77 \\
10             &  19.08\% &  10.78\% &  43.23\% &      4.32 \\
11             &  26.95\% &  17.94\% &  47.67\% &      5.24 \\
12             &  21.33\% &  11.32\% &  47.76\% &      5.73 \\
13             &  20.89\% &  12.98\% &  47.40\% &      6.16 \\
14             &  16.32\% &   6.27\% &  46.29\% &      6.48 \\
15             &  14.99\% &   6.14\% &  45.53\% &      6.83 \\
16             &  25.05\% &  15.15\% &  54.22\% &      8.67 \\
17             &  20.74\% &   9.61\% &  56.95\% &      9.68 \\
18             &  16.78\% &   6.44\% &  52.27\% &      9.41 \\
19             &  18.41\% &   7.51\% &  49.47\% &      9.40 \\
20             &  24.64\% &  14.60\% &  59.33\% &     11.87 \\
\bottomrule
\end{tabular}
\caption{Percentage reduction in instance size ($n$, $m$) and number of cliques ($k$) after preprocessing. TF and LV data are combined and grouped by initial $k$ value (prior to preprocessing). Average (absolute) number of cliques removed is also reported.}
\label{table:preproc}
\end{table}

\AlgLp and \AlgIp both have runtime exponential in the number of cliques ($k$).
Pruning away easily detectible cliques before running the expensive decomposition algorithms reduces the size of the input parameter, thus reducing the overall runtime.

The preprocessing works by running a modified breadth-first search algorithm to detect and remove cliques that are either (1) disjoint from the rest of the network (i.e. in their own connected component) or (2) intersect other cliques exclusively on single vertices (i.e., share no edges with other cliques). In any valid decomposition, a clique $C$ of size $\ell$ with edges of weight $w$
from either of these categories must be represented by between one and $\ell(\ell-1)/2$ cliques, all with weight $w$. Thus, if
there is any solution with $k$ cliques, there must also be a solution which includes $C$ (with weight $w$) and has at most $k$ cliques. Thus, removing $C$ from the instance and reducing $k$ by 1 will not change whether or not we have a yes-instance (and $C$ can be added to the resulting set to form a valid solution for the original instance). Because our algorithm depends exponentially on $k$, this pre-processing results in much faster overall runtimes.

Our algorithm first loops over each $v_i \in V$ checking if all its incident edges have the same weight. If they do, call this edge weight $w_i$ and $U_i$ the set of vertices adjacent to $v_i$.
We then iterate over each $u_j \in U_i$ and check that there exists an edge from $u_j$ to $u_k$ for
each $u_k \in U_i$ with edge weight equal to $w_i$.
If this process returns true, then the set $U_i \cup v_i$ forms a clique. Since all edge weights are equal, the clique does not share any edges with other cliques, making it safe for removal. This process takes $O(VE)$ time.
In our experiments, preprocessing took an average of $4.67$ seconds to run on the TF graphs, and $33.71$ seconds on the LV graphs. Table ~\ref{table:preproc} shows the average reduction in instance sizes and clique counts.

\bibliographystyle{plainurl}
\bibliography{references}

\end{document}